\documentclass[reprint,superscriptaddress, amsmath,amssymb, aps, pra, longbibliography]{revtex4-1}

\usepackage{natbib}
\usepackage{soul}
\usepackage{amsmath}
\usepackage{amssymb}
\usepackage{graphicx}
\usepackage{dcolumn}
\usepackage{bm}
\usepackage{amsfonts}
\usepackage{subfigure}
\usepackage{array}
\usepackage{float}
\usepackage{color}
\usepackage[colorlinks=true,linkcolor=blue]{hyperref}%
\hypersetup{allcolors=blue}
\usepackage[normalem]{ulem}
\usepackage{xcolor}

\newcommand{\final}{\text{final}}
\newcommand{\init}{\text{init}}

\newcommand{\ket}[1]{\ensuremath{\left\vert #1 \right\rangle}}
\newcommand{\quantmean}[1]{\ensuremath{\left\langle #1 \right\rangle}}

\newcommand{\tgt}{\text{tgt}}

\begin{document}

\title{Principles of tractor atom interferometry}
\date{\today }

\author{Georg Raithel}
\affiliation{Department of Physics, University of Michigan, Ann Arbor, MI 48109}
\author{Alisher Duspayev}
\email{alisherd@umich.edu}
\affiliation{Department of Physics, University of Michigan, Ann Arbor, MI 48109}
\author{Bineet Dash}
\affiliation{Department of Physics, University of Michigan, Ann Arbor, MI 48109}
\author{Sebasti\'an C. Carrasco}
\affiliation{DEVCOM Army Research Laboratory, 2800 Powder Mill Road, Adelphi, MD 20783}
\author{Michael H. Goerz}
\affiliation{DEVCOM Army Research Laboratory, 2800 Powder Mill Road, Adelphi, MD 20783}
\author{Vladan Vuleti\'c}
\affiliation{Department of Physics, MIT-Harvard Center for Ultracold Atoms, and Research Laboratory of Electronics, Massachusetts Institute of Technology, Cambridge, MA 02139}
\author{Vladimir S. Malinovsky}
\affiliation{DEVCOM Army Research Laboratory, 2800 Powder Mill Road, Adelphi, MD 20783}

\begin{abstract}
We present possible design concepts for a tractor atom interferometer (TAI) based on three-dimensional confinement and transport of ultracold atoms. The confinement reduces device size and wave-packet dispersion, enables arbitrary holding times, and facilitates control to create complex trajectories that allow for optimization to cancel unwanted sensitivity, fast splitting and recombination, and suppression of detrimental nonadiabatic excitation. Thus, the design allows for further advancement of compact, high-sensitivity, quantum sensing technology. In particular, we focus on the implementation of quantum-enhanced accelerometers and gyroscopes. We discuss TAI protocols for both spin-dependent and scalar trapping potentials. Using optimal control theory, we demonstrate the splitting of the wave function on a time scale two orders of magnitude shorter than the previous proposal using adiabatic dynamics, thus maximizing the time spent at full separation, where the interferometric phase is accumulated. Lastly, we explore the possibility of including non-classical correlations between the atoms to improve sensitivity. The performance estimates for TAI give a promising perspective for atom-interferometry-based sensing, significantly exceeding the sensitivities of current state-of-the-art devices.
\end{abstract}

\maketitle

\section{Introduction}
\label{sec:intro}

Since their first demonstrations
\cite{Carnal1991,Keith1991,Riehle1991,Kasevich1991}, atom interferometers~\cite{Cronin2009,Berman1997,Abend2020}  have become a powerful tool with a broad range of applications in fundamental physics, {\emph{e.g.}}, testing the equivalence principle, free fall and (non)-Newtonian forces~\cite{Tarallo2014,Schlippert2014,Kovachy2015,Jaffe2017,Rosi2017,Fixler2007,Menoret2018,Xu2019}, 
gravitational-wave detection~\cite{Hogan2016}, precision measurements of atomic constants~\cite{Hanneke2008,Parker2018,Morel2020} and applied science, \emph{e.g.}, inertial 
sensing~\cite{Barrett2019,Bongs2019,Alzar2019}
and geodesy~\cite{Bongs2019, Bidel2020,Tino2019}.
Previous work on AI includes 
free-space~\cite{Asenbaum2017,Rosi2014,Hamilton2015}
and point-source~\cite{Dickerson2013,Hoth2016,Chen2019} 
AI, as well as guided-wave AI experiments~\cite{Wu2007,Moan2020,Sapiro2009} 
and proposals~\cite{Davis2008,Zimmermann2019}.
Free-space and point-source AIs typically employ atomic fountains or dropped / freely expanding atom clouds. The point-source method supports efficient readout and data reduction~\cite{Chen2020}, 
enables high bandwidth, and affords efficiency in the partial-fringe regime. Atomic fountains, typically employed in free-space AI, maximize interferometric time and thus increase 
sensitivity~\cite{Asenbaum2017,Rosi2014,Hamilton2015}, but may require large experimental setups. Guided-wave AIs offer compactness and are often used as Sagnac rotation sensors, but are susceptible to noise in the guiding potentials. In both free-space and guided-wave AI, wave-packet dynamics along unconfined degrees of freedom can cause wave-packet dispersion and failure of the split wave packets to recombine. Coherent recombination of split atomic wave functions upon their preparation and time-evolution remains challenging in recent AI studies~\cite{Stickney2002,Stickney2003,Burke2008,Stickney2008}.
Atom interferometry is a cornerstone of space-based fundamental and applied research in the cold-atom lab (CAL~\cite{Frye2021}), where decoherence due to guide- and trap-induced forces and apparatus-size issues, otherwise encountered due to free fall, are significantly reduced. Wave-packet dispersion and atomic interactions as well as practical problems associated with efficient closure control still remain even at CAL and its successors. 

Here, we describe tractor atom interferometry (TAI), a method based upon uninterrupted three-dimensional (3D) confinement and guiding throughout the AI sequence. The paper is organized as follows. In Sec.~\ref{sec:concept} the TAI concept and its key features are explained. 
In Sec.~\ref{sec:lattice} we discuss an implementation with spin-dependent optical lattices of $^{87}$Rb on the D1 line ($5S_{1/2} \leftrightarrow $ $5P_{1/2}$) with $\pi/2$-splitting and recombination pulses driven by a Raman transition. 
While spin-dependent lattices afford robust AI COM mode splitting~\cite{Duspayev2021}, they require multiple lattice laser beams and have other drawbacks, explained in Sec.~\ref{sec:lattice}. In Sec.~\ref{sec:scalar} we discuss a method of TAI on a scalar potential on which rapid COM mode splitting is achieved by a quantum control method. Conclusions and impacts of TAI are discussed in Sec.~\ref{sec:discussion}

\section{Concept}
\label{sec:concept}

TAI differs from cold-atom free-space, point-source, and guided-wave AI in that the interfering atomic wave-packet components are transported in conservative, sub-micron to mm-sized, 3D traps that are formed by tractor potentials that move on predetermined trajectories~\cite{hanselpra, steffen2012, Duspayev2021}.  The traps can be implemented via optical tweezers (tractor beams)~\cite{Endres1024, Barredo1021, kim2016, ohl}, optical lattices~\cite{kovachypra2010, Kumar2018, klostermann, krinner2018}, RF-dressed potentials~\cite{white2006, Colombe_2004, V_Guarrera_2015, Garraway_2016, Navez_2016}, optical or magnetic potentials on atom chips~\cite{keil2016, berrada, berrada2016}, etc., and any combination of these~\cite{bayha2020, deist2022, trisnadi2022}. Uninterrupted 3D confinement in tractor traps (1) guarantees recombination, (2) allows arbitrary holding times, directional reversal, complex trajectory patterns for cancellation of sensitivities to inertial forces that are not of interest, and (3) addresses signal degradation caused by wave-function dispersion and limitations in recombination control. Ideally, the AI wave-function components are given by the 3D vibrational ground states of the tractor traps at all times during the AI sequence. As we will show in this paper, this condition may be relaxed in order to realize fast AI splitting and recombination with coherent quantum control methods.

In TAI, the tractor controls (laser-beam angles, diameters, powers and phases, electric and magnetic fields) define  pre-determined trajectories,
${\bf{x}}_m(t)$, of the tractor-potential minima in 3D configuration space. 
The trajectories mark the centers of the tractor traps versus time.
A pair of traps, indexed by $m=1, 2$, are intersecting at initial and final space-time points, denoted
${\bf{x}}_{\init}(t_{\init})$ and ${\bf{x}}_{\final}(t_{\final})$, respectively. This situation, while classically forbidden due to the uniqueness of classical trajectories, can be realized in quantum mechanics by employing a pair of spin states with different, state-specific tractor traps on spin-dependent potentials that coincide at the initial and final space-time points, or by AI splitting and recombination afforded by quantum tunneling or some other type of coherent dynamics between a pair of potential wells on a spin-less (scalar) potential. In these two cases, the AI beam-splitters and re-combiners are implemented via microwave or Raman laser pulses that couple the active spin states, or by quantum manipulation on a dynamic double-well landscape, respectively.

Following the usual path integral formalism, the interferometric phase of the TAI is given by $\Delta \Phi = \frac{S_2 - S_1}{\hbar}$, with the actions computed as,
\begin{equation}
S_m = \int_{t_{\init}}^{t_{\final}} \mathcal{L} \big( {\bf{x}}_m (t),\dot{\bf{x}}_m (t),t \big) dt \quad .
\label{eq:action}
\end{equation}
where $\mathcal{L}$ is the Lagrangian function for a trajectory ${\bf{x}}_m(t)$~\cite{Cronin2009}. In TAI, the latter is given by the pre-programmed center locations of the tractor traps~\cite{Duspayev2021}, so that Eq.~(\ref{eq:action}) can be evaluated directly, without having to perform a classical trajectory calculation first. This contrasts with AI-types that have one or more generalized classical degrees of freedom. In those cases, the classical trajectories ${\bf{x}}_m(t)$ are not {\emph{a priori}} known and must be computed {\emph{before}} Eq.~(\ref{eq:action}) can be evaluated. It is implicit to the TAI method that automatic closure of the interferometer can be guaranteed via correct tractor programming.

The dependencies of the differential interferometric phases on rotation and acceleration scale as~\cite{Cronin2009}
\begin{eqnarray}
\Delta \Phi_{\Omega} & =  & \frac{2 m K {\bf{\Omega}} \cdot  {\bf{A}}}{\hbar} \nonumber \\
\Delta \Phi_{a} & \approx  & \frac{m K a z T}{\hbar}  \quad .
\label{eq:phases}
\end{eqnarray}     
Here, $m$ is the atom mass, $a$ the acceleration, ${\bf{A}}$ the interferometric area, ${\bf{\Omega}}$ the frame's angular velocity measured against an inertial frame, $K$ the number of loops in the TAI sequence, $z$ the well separation along the acceleration vector, and $T$ is the AI time. The expression for the acceleration phase given in Eq.~(\ref{eq:phases}) is only approximate, because details of how the TAI wells separate and recombine play a role. The acceleration phase can be calculated accurately after the exact tractor trajectories ${\bf{x}}_m(t)$ have been specified.
A discussion of quantum-projection-limited sensitivity levels for rotation and acceleration is provided in~\cite{Duspayev2021}. Phase sensitivity estimates for some conditions of the presently discussed schemes are given in Sec.~\ref{sec:discussion}.

\begin{figure*}[tbp]
\begin{center}
\includegraphics[width=0.93\textwidth]{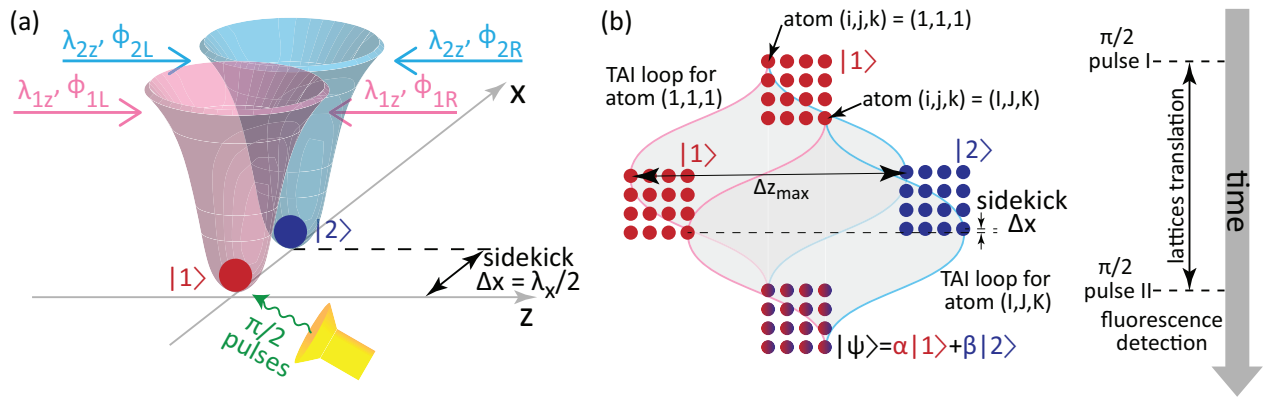}
\end{center}
\caption{Concept of TAI in a spin-dependent lattice. (a) Two different spin states of the ground-state atom, $\ket{1}$ and $\ket{2}$, which are here taken to be the $\ket{F = 1, m = 0}$ and $\ket{F = 2, m = 0}$ levels of $^{87}$Rb, are trapped in respective 3D optical lattices. For simplicity, single atoms in respective individual lattice wells are shown. The confinement of the lattice wells in the $y$-direction is static and is, for clarity, suppressed in the figure. The optical lattice in $x$-direction features a short-distance differential ``sidekick" of half the lattice period along $x$, $\Delta x = \lambda_{x} / 2$. The sidekick suppresses collisions between the spin states during the TAI translation sequence, which mostly consists of a long-distance lattice translation in $z$-direction, shown in panel (b) as a function of time. The total sequence includes $\pi/2$ AC-shift-free microwave or Raman pulses to open and close the interferometer. The final AI signal is acquired via fluorescence imaging readout after completion of TAI loops. See text for more details.}
\label{Fig1}
\end{figure*}

TAI differs from other work on cold-atom free-space, point-source, and guided-wave implementations of AI (see Sec.~\ref{sec:intro}) in that the interfering wave-function components are confined in 3D at all times, suppressing dispersion and allowing for maximum control. Proper programming of the tractor traps ensures AI closure. The robustness of TAI closure against tractor-induced and background inertial effects is limited by the forces of constraint acting on the atoms. 
The forces of constraint are given by the forces that the tractor potentials exert on the trapped atoms to keep them on the pre-programmed tractor trajectories, ${\bf{x}}_m(t)$. The forces of constraint counter-balance the inertial forces, $m \ddot{{\bf{x}}}_m(t)$, in the instrument's frame of reference, as well as the 
inertial forces caused by the motion of the platform the instrument is mounted on.  
Since optical-lattice traps, considered here, can exert forces of constraint that exceed gravity on Earth by orders of magnitude,
closure in optical-lattice-based TAI can be very robust. As a result, optical-lattice-based TAI may be implemented in scenarios that require large forces of constraint.
Uninterrupted 3D confinement of the atomic wave-function components in the tractor traps further eliminates uncontrolled wave-packet dispersion.
Geometry and speeds of the TAI tractor trajectories are user-programmable and flexible, including multi-loop designs, trap-hold intervals, and twisted patterns. Hence, TAI can be adapted to a variety of applications. 

The TAI concept translates well to microgravity implementations, where the tractor-trap depth can be relaxed into the sub-Hz regime at times when the forces of constraint become very small. Trap relaxation efficiently addresses concerns with phase noise in Eq.~(\ref{eq:phases}) caused by trap-depth fluctuations. In relaxed tractor traps, the AI time, $T$, may extend to minutes, which translates into greatly enhanced sensitivities. Also, under such conditions the motional time scale of the atoms in the tractors becomes so slow that technical noise in the acoustic and higher-frequency bands does not couple to the vibrational dynamics of the atoms.

\section{TAI in spin-dependent optical lattice}
\label{sec:lattice}

One implementation of TAI is based on the use of spin-dependent optical lattices. 
Fig.~\ref{Fig1} outlines the concept. The lattice spin states are the $\vert 1 \rangle :=$ 
$\vert F=1, m=0 \rangle$ [red in
Fig.~\ref{Fig1}~(a)] and $\vert 2 \rangle := $ 
$\vert F=2, m=0 \rangle$ (blue) levels of $^{87}$Rb, which are magnetic-field-insensitive in lowest order. The states are trapped in respective 3D optical lattices with spatial periodicities $\lambda_{m,n}$, where the first index refers to the ket $\vert m \rangle$ and the second to the spatial axes, $n = x$, $y$ and $z$. For lattices formed by counter-propagating beam pairs the periodicities are given by half the optical wavelength ($\lambda_{m, opt} \approx 795$~nm for the Rb D1 line). The spatial peridiodicities may be increased by choosing beam-pair angles $\theta_{m,n} \le \pi$, for which $\lambda_{m,n}=$ $ \lambda_{m, opt}/ [2 \sin(\theta_{m,n}/2)]$. It must be ensured that $\lambda_{1,n} = \lambda_{2,n} = : \lambda_{n}$ for all $n=x$, $y$ and $z$. The lattices for the two spin states can be translated relative to each other using independently controlled phases, $\phi_{m,L}$ and  $\phi_{m,R}$ with $m=1, 2$ and $L$ = ``left'' and $R$ = ``right'' (index $n$ is suppressed for brevity).  

An implementation of TAI may proceed as follows. The lattice structures along $y$ are static and are overlapped at all times. A differential ``sidekick'' between the lattices for the $\vert 1 \rangle$ and $\vert 2 \rangle$ atoms displaces the respective lattice-trapped wave-function components relative to each other in the $x$-direction by a distance $\Delta x = \lambda_{m,x}/2$, {\emph{i. e.}} half the spatial lattice period along $x$. In that way, 
the subsequent long-distance tractor motion along $z$ will not lead to collisions between the $\vert 1 \rangle$ and $\vert 2\rangle$ atoms. 
As shown in Fig.~\ref{Fig1}~(b), the $z$-translations form a large AI area in the space-time plane, which is  
suitable for inertial sensing (gray areas). The lattice has $I \times J \times K$ sites, with $I$, $J$, and $K$ denoting the number of sites in the $x$, $y$ and $z$-directions, and integer indices $(i, j, k)$ labeling individual sites. Fig.~\ref{Fig1}~(b) shows the TAI trajectories for the corner sites $(1,1,1)$ and $(I, J, K)$. The AI sequence involves $\pi/2$ AC-shift-free microwave or Raman pulses to open and close the interferometer. An AI signal can be acquired via imaging of spin-dependent fluorescence after completion of the TAI loops.

We note that the actual laser-beam frequencies for trapping along $x$, $y$ and $z$ have to differ by an amount significantly larger than the vibrational frequencies of the atoms in the lattice traps to exclude any effects of optical interference beats on the center-of-mass (vibrational) wave functions of the trapped atoms. It is expected that this condition will be 
satisfied in most practical implementations of optical-lattice-based TAI.

\begin{figure*}[tbp]
\begin{center}
\includegraphics[width=0.98\textwidth]{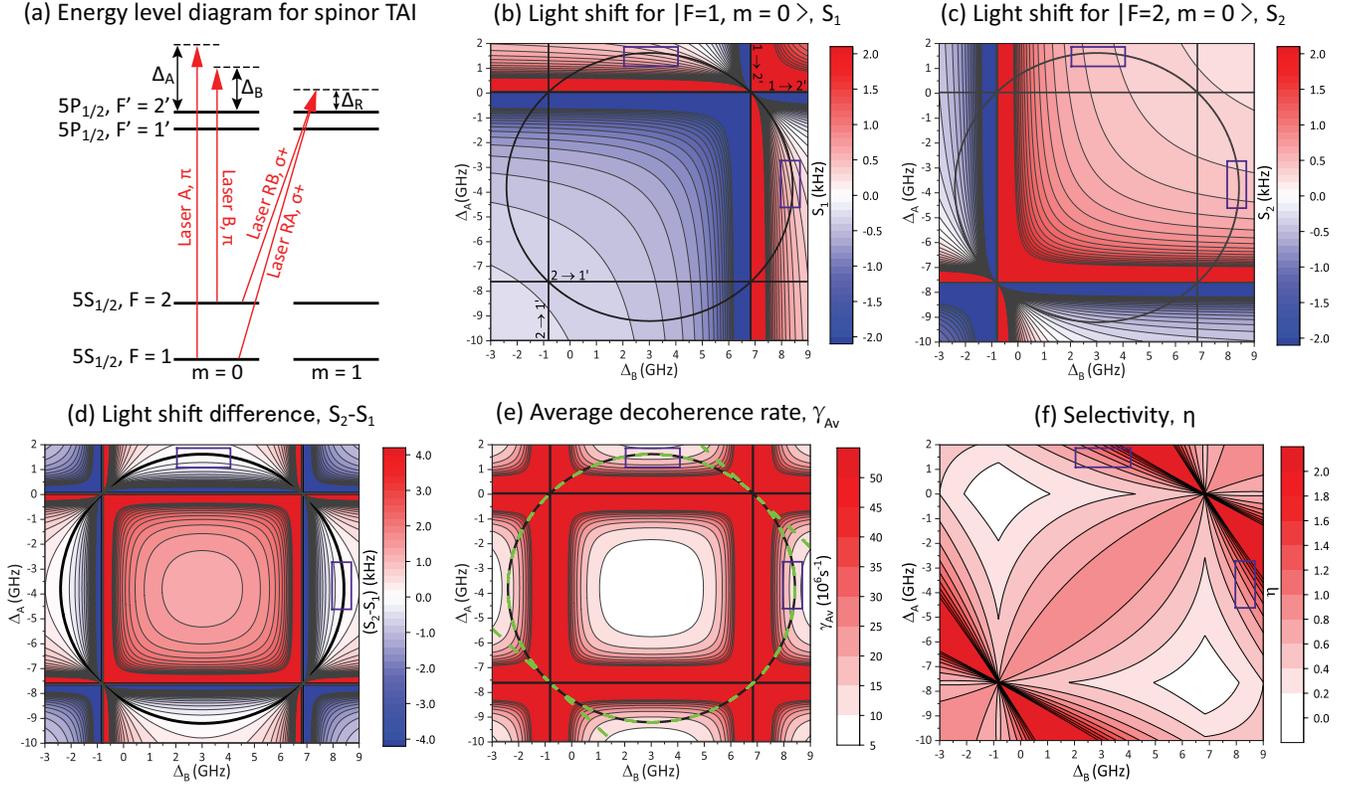}
\end{center}
\caption{Panel (a) shows spin states, trapping-field configurations and detuning definitions for a case of 
magnetic-field-insensitive spinor TAI in $^{87}$Rb. Panels (b) and (c) show the light shifts, $S_1$ and $S_2$, of states $\vert 1 \rangle$ and $\vert 2 \rangle$ on the $(\Delta_A,\Delta_B)$-plane, respectively, and panel (d) shows the difference $S_2-S_1$.
Panel (e) shows the average photon scattering rate (decoherence rate), $\gamma_{Av}$, on the $(\Delta_A,\Delta_B)$-plane, and panel (f) the selectivity, $\eta$, as defined in the text, of the $A$- and $B$-fields in selectively trapping atoms in states  $\vert 1 \rangle$ and $\vert 2 \rangle$.}
\label{Fig2}
\end{figure*}

We first discuss the case of spin-dependent lattices formed with a pair of near-resonant laser fields near the D1 line of Rb (wavelength $\lambda_{m,opt} \approx 795$~nm for $m=1$ and $2$). The two fields are labeled $A$ and $B$ [see Fig.~\ref{Fig2}~(a)]. Each one of the indicated  magnetic-field-insensitive spin states, $\vert 1 \rangle$ and $\vert 2 \rangle$,  exhibits two light shifts from fields $A$ and $B$ that we label $S_{m,A}$ and $S_{m,B}$, with $m=1,2$. The net shift for each level then is $S_m=S_{m,A} + S_{m,B}$. All shifts depend on atom-field detuning. For specificity, we measure the frequency offset of the $A$-field relative to the $F=1$ to $F'=2$ hyperfine transition of the $^{87}$Rb D1 line, and that of the $B$-field relative to the $F=2$ to $F'=2$ transition [see Fig.~\ref{Fig2}~(a)]. The respective frequency detunings are denoted $\Delta_A$ for the $A$- and $\Delta_B$ for the $B$-field. For a sample case of equal electric-field amplitudes of $E_A=E_B = 100$~V/m, in Figs.~\ref{Fig2}~(b) and (c) we show the level shifts for the spin states $\vert 1 \rangle$ and $\vert 2 \rangle$, respectively, and in Fig.~\ref{Fig2}~(d) the difference, $S_2 - S_1$.

In addition to reducing phase fluctuations caused by differential trap-beam intensity noise, it is imperative to reduce coherence loss due to photon scattering of trap-beam light. Coherently-split wave-function components as sketched in Fig.~\ref{Fig1} are susceptible to this type of coherence loss. As a quantitative figure, we use the photon scattering rate of a coherently split atom averaged over both internal spin states. Denoting the scattering rate of the atom in pure spin state $\vert m \rangle$ due to fields $A$ and $B$ as
$\gamma_{m,A}$ and $\gamma_{m,B}$, with $m=1,2$, respectively, the scattering rate in spin state $\vert m \rangle$ is $\gamma_m = \gamma_{m,A} + \gamma_{m,B}$. Assuming that the atomic wave packets are split evenly between the spin states, the average scattering rate is $\gamma_{Av} = (\gamma_1 + \gamma_2)/2$, which is shown in Fig.~\ref{Fig2}~(e).
Unsurprisingly, it is seen in Fig.~\ref{Fig2}~(e) that the overall scattering is minimized if both detunings $\Delta_A$ and $\Delta_B$ are about equally far away from the nearest hyperfine resonances. Since the light shifts, their difference, and the photon scattering rates all scale linearly in field intensity, the results in Fig.~\ref{Fig2} easily scale to configurations with different (but equal) fields,  $E_A=E_B$. Slight modification of the calculation also allows one to consider cases with $E_A \ne E_B$.

\begin{figure}[btb]
\includegraphics[width=0.48\textwidth]{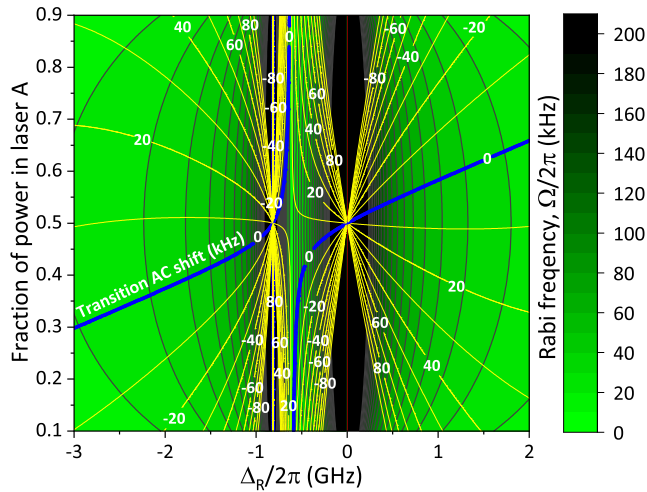}
\caption{Raman Rabi frequency (colored scale) and AC shifts (contour lines with labels) 
due to Raman beams near the D1 line for the transition 
between the states $\ket{1}$ and $\ket{2}$ vs intermediate detuning $\Delta_R$ shown in Fig.~\ref{Fig2}~(a) and the fraction of power in the $A$-beam (that connects to $\ket{1}$). 
Both fields are $\sigma^+$-polarized relative to the quantization axis that defines the atomic spin states 
$\ket{1}$ and $\ket{2}$. In this example, the total power of the Raman coupling beams $A$ and $B$ is $P_{total}$ = 100~mW, and the assumed Gaussian beam parameter is $w_0 = $1~cm.} 
\label{Fig3}
\end{figure}

Inspection of Fig.~\ref{Fig2} shows that detuning combinations in the $(\Delta_A,\Delta_B)$-plane that minimize photon scattering 
under the condition $S_1 = S_2 \approx 400$~Hz have $\gamma_{Av} \approx 12$~Hz, so that $S_m / \gamma_{Av} \approx 40$. It becomes obvious that TAI with spin-dependent lattices will require blue-shifted configurations, in which the light-shift traps localize the atoms at locations of minimal field intensity, hence minimizing the actual photon scattering rate (decoherence rate). 
This corresponds with regions within the 
blue rectangles in Fig.~\ref{Fig2}. At the field minima in low-field-seeking lattice traps one may then expect effective coherence loss rates on the order of $\sim 1/500$ of the trap depth. While this may suffice for demonstrations, high-precision AI with long AI times $T$ will require schemes with a lower coherence loss rate during times at which the wave-function components are held at a large mutual separation. This may be achievable by combinations of laser-intensity sweeps and laser-frequency sweeps farther away from resonance. The sweeps enable a reduction of the decoherence rate at times when the wave-function components are spatially separated. Effectiveness of such methods will require that the wave-function components are spatially separated during most of the AI time $T$ (which is, typically, the case).     

Another figure of merit that applies, in particular during the splitting phase of TAI, is the degree to which the optical potentials are spin-dependent. 
For the survey presented here, we compute the spin-selectivity figure $\eta_m  =  (\pm S_{m,A} \mp S_{m,B}) / (S_{m,A} + S_{m,B})$, where the upper signs apply to state $\vert 1 \rangle$ and the lower ones to $\vert 2 \rangle$. In an ideal scenario of prefect spin selectivity, field A would only trap state $\vert 1 \rangle$ and not induce AC shifts in state $\vert 2 \rangle$, and similarly field B would only trap state $\vert 2 \rangle$ and not induce AC shifts in state $\vert 1 \rangle$. In that case, it would be $\eta_1= \eta_2 = 1$. For a comprehensive reading, in Fig.~\ref{Fig2}~(f) we only show $\eta = (|\eta_1-1| +  |\eta_2 -1|)/2$. Ideal selectivity will then require $\eta = 0$, and the more $\eta$ exceeds the value of 1 the more marginal the spin selectivity will become.  The results for $\eta$, presented in Fig.~\ref{Fig2}~(f), show that ideal spin selectivity only occurs in regions on the $(\Delta_A,\Delta_B)$-plane where the decoherence rate is large. A compromise between reasonably low decoherence rates and high spin selectivity of the traps occurs near the left margin of the upper blue rectangle and near the lower margin of the lower blue rectangle in Fig.~\ref{Fig2}.

We next consider the $\pi/2$ beam-splitting in spinor TAI for opening and closing the interferometer. For the case considered here, the transition $\vert 1 \rangle $ $\rightarrow$ $\vert 2 \rangle$ can be coherently driven with an RF field at the ground-state hyperfine splitting frequency of $6.8$~GHz or via an optical Raman transition. Since the pulse time must be much smaller than the AI time $T$, we envision coupling Rabi frequency, $\Omega$, in the range of 50~kHz. These will enable $\pi/2$-pulses with durations $\lesssim 0.01$~ms even for cases of short TAI sequences with $T \sim 1$~ms, as will be required in initial testing of TAI. For specificity, here we consider an optical Raman coupling with field polarizations as shown in Fig.~\ref{Fig2}~(a). The Raman transition has two variables, the Raman detuning, $\Delta_R$, and the power splitting ratio between the two fields driving the transition. We seek a configuration that results in a decent $\Omega$, while minimizing the light shift due to the Raman-transition fields. In Fig.~\ref{Fig3} we show $\Omega$ vs $\Delta_R$ and the fraction of power in the $A$-beam (which connects to the $\vert 1 \rangle$-state). The overlay shows the AC shift induced by the Raman beams. For the anticipated pulse durations of $\sim 0.01~$ms, $\Omega$ near 50~kHz and an AC shift of $\lesssim 10$~kHz would be desirable. Fig.~\ref{Fig3} shows that such conditions can be attained with $\Delta_R \approx 1.5$~GHz and about $2/3$ of the power in the $A$-beam, for the net laser power and beam size as specified in the caption.

Figs.~\ref{Fig2} and~\ref{Fig3} can be summarized as follows. It is possible to generate spin-dependent lattices for pairs of magnetic-field-insensitive hyperfine ground states of alkali atoms using two lattice laser fields with different frequencies. The lattices can be ``magic'', in a sense that both spin states experience the same trapping potentials for phases $\phi_{m,L}$ and  $\phi_{m,R}$ in Fig.~\ref{Fig1} at which the spin-dependent potentials are overlapped. The magic condition simplifies efficient $\pi/2$ AI splitting between the vibrational ground levels of the spin states $\vert 1 \rangle$ and
$\vert 2 \rangle$ in their respective lattices, and it reduces AI phase noise in $\Phi$ (see Eq.~\ref{eq:phases}) caused by laser intensity fluctuations. Decoherence rates due to photon scattering limit the AI time, $T$, to values on the order of 500 times the inverse of the trap depth (in Hz). Parameters that yield sufficient spin selectivity of the traps as well as AC-shift-free $\pi/2$ Raman pulses for opening and closing the interferometer exist. For spinor-TAI with long interferometer times $T$, required to reach high sensitivities for inertial fields, it will be necessary to use combinations of laser-intensity sweeps and laser-frequency sweeps to intermittently reduce decoherence rates due to photon scattering at times when the wave-function components are spatially separated, in a manner that 3D trapping at all times remains guaranteed. 
In this scenario, the lattice tractor traps become spin-independent at the times when the spin components are separated, because the trap laser fields will be temporarily swept far-off-resonance to reduce the photon scattering. 

While we believe that the spinor TAI method can the realized in the laboratory, it also is prudent to consider scalar TAI~\cite{Duspayev2021} as an alternative method. 
In scalar TAI, the laser fields can be held far-off-resonance at all times, allowing one to reduce decoherence caused by photon scattering to practically zero. In scalar TAI, the trapping laser will be far-off-resonant, allowing the use of high-power, highly efficient lasers (for instance, YAG lasers or frequency-doubled telecom lasers). There will be only a single trapping potential, on which single wells are split into double wells. These follow different tractor trajectories and are recombined back into single wells at AI closing. The challenge in that scheme is to provide fast splitting without exciting the split wave packets into higher vibrational states of the atoms in the tractor traps~\cite{Duspayev2021}. In the following section we will present a method of scalar TAI that employs rapid beam splitters that are based on quantum-state engineering methods.

\section{Scalar TAI with quantum-state-engineering-enabled beam splitters}
\label{sec:scalar}

In the spin-less (or scalar) case, the goal is to split the wave function into a superposition of each tractor trap ground state as quickly as possible. This maximizes the available time for accumulating a differential phase.
Mathematically, the goal is to find the tractor functions ${\bf{x}}_m(t)$ that minimize the functional
\begin{equation}
    J[{\bf{x}}_m (t)] = |\quantmean{\Psi_\tgt | \Psi(T_\text{s})}|^2 \ ,
\end{equation}
where $\ket{\Psi_\tgt} = (\ket{\Psi_\text{L}} + \ket{\Psi_\text{R}}) / \sqrt{2}$ is the target state, $\ket{\Psi_{\text{L}, \text{R}}}$ are the ground states of each trapping potential when separated at a distance $\Delta z$, and $\ket{\Psi(T_\text{s})}$ is the system state after allowing a time $T_\text{s}$ for splitting. The functional corresponds to the fidelity at the end of the splitting. To preserve the symmetry between the arms, we assume ${\bf{x}}_1(t) = - {\bf{x}}_2(t) = {\bf{x}}(t)$. In this way, we only need to find a single tractor function. Furthermore, we assume that the trapping potential only allows movement in the $z$ direction so that the dynamics can be described as ${\bf x}(t) = z(t) \, \hat z$, where $\hat{z}$ is the unit vector in the $z$ direction. Consequently, the equation of motion is
\begin{equation}
\begin{gathered}
  i \hbar \frac{\partial}{\partial t} \Psi(z, t) =  H(z,t) \Psi(z, t)\,,\quad\text{with}\\
  H(z,t) = - \frac{\hbar^2}{2 m} \frac{\partial^2}{\partial z^2} + V(z - z(t)) + V(z + z(t))\,,
\end{gathered}
\label{Schroedinger}
\end{equation}
where $V(z) = - V_0 \exp \left[ - 2 \log(2) z^2 / d^2 \right]$ is the trapping potential. For a specific example, we use the parameters $V_0 = 0.5$ MHz for the potential depth, $d = 23.5$ $\mu$m for the full width at half maximum, and the mass of the $^{87}$Rb atom for $m$. These parameters equal those used in Ref.~\cite{Duspayev2021}.

In Ref.~\cite{Duspayev2021}, the tractor function $z(t)$ was tuned adiabatically, so that $\Psi(z, t)$ is the ground state of the shifted $V(z(t))$ at all times. The time scale for achieving an adiabatic separation is on the order of seconds, that is, comparable to the accumulation time when the wave packets remain at maximum separation. When $z(t)$ changes faster than adiabatically in order to minimize $T_\text{s}$ and maximize the accumulation time, the movement can produce unwanted excitations and even cause the particles to escape the trap. We will demonstrate that coherent quantum control can refocus the dynamics to remove these nonadiabatic effects, thus achieving fast splitting.

We will use the following approach for coherent quantum control. Given a value of $T_\text{s}$, we set the $x$ and $y$ components of the tractor function to zero and specify the $z$ component as a piecewise linear function with $n$ equally spaced segments. Therefore, the tractor function is fully described by $n-1$ parameters, \emph{i.e.}, the values of the tractor function $x_j$ at the carrier points $t_j$. The values at $t=0$ and $t=T_\text{s}$ are fixed at 0 and $\Delta z$, respectively. The piecewise linear ansatz reduces the dimensionality of the parameter space sufficiently to find a ``guess'' tractor function with non-zero fidelity by testing random seeds.

Once we find an appropriate guess solution, we can calculate the gradient $\partial J / \partial z_j$ using central finite differences and use it to feed a quasi-Newton gradient-based optimizer such as L-BFGS-B~\cite{2020SciPy-NMeth}. 
Each evaluation of $J$ runs in parallel using a split-operator scheme to propagate the wave function. If further refinement is required, we can double the number $n$ of segments, matching the lower-dimensional solution, and repeat the optimization. Another advantage of this approach is that it decouples the number of control parameters from the number of time steps. This is useful when the time scale of the dynamics inside the trap is much faster than the splitting timescale.

\begin{figure}[tbp]
\centering
\includegraphics[width=0.48\textwidth]{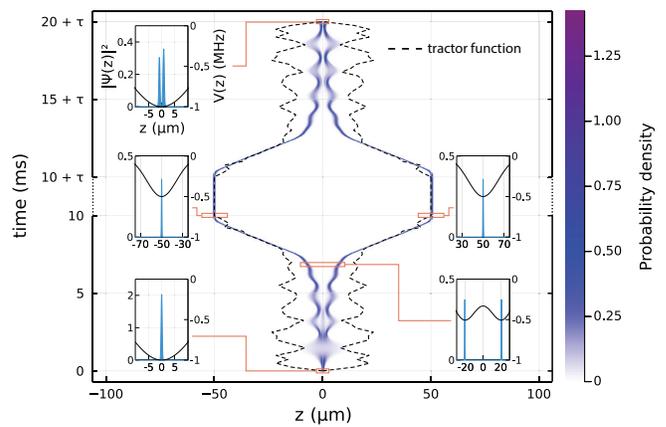}
\caption{ Optimized interferometric sequence for scalar TAI. The heatmap shows the probability density $|\Psi(z, t)|^2$ as the initial wave packet passes through splitting (first $10$~ms), free evolution (not to scale), and recombination (last $10$~ms). The dashed lines show the tractor functions, \emph{i.e.}, the optimized controls. The insets correspond to selected snapshots of the dynamics showing both the potentials (black lines) and the probability densities (blue lines).} 
\label{Fig4}
\end{figure}

Fig.~\ref{Fig4} illustrates the entire scalar TAI scheme using the coherent quantum control solution, including both splitting and recombination. The heatmap represents the probability density. The dashed lines show the tractor function, \emph{i.e.}, the minima of the trapping potentials. The insets show the probability density and the overall potential (the sum of both trapping potentials) at different time instances. When calculating the tractor function, we start with $10$ segments, corresponding to 11 carrier points distributed over a splitting time of $T_\text{s} = 10$~ms, 
and obtain a fidelity of $0.826$. We then increase the number of segments to $80$ (81 carrier points distributed over the same $T_\text{s}$), 
to obtain a fidelity of $0.991$. This splitting time represents an improvement over the original spin-less TAI result~\cite{Duspayev2021} by two orders of magnitude. As shown in Fig.~\ref{Fig4}, the tractor function does not simply separate both trapping potentials to split the wave function but oscillates to refocus it and avoid unwanted excitations. After that, the wave function starts to travel with the minima of the trapping potential to the maximum separation (see the bottom right inset). Before reaching maximum separation, the tractor function suddenly reduces the separation to decelerate the wave packets, so they remain in the ground state for a time $\tau$ to accumulate the interferometric phase (see middle insets). Note that the accumulation time is typically several orders of magnitude longer than the splitting time. Finally, the recombination repeats the splitting process, but running backward. In the end, we obtain a state that is a linear combination of the ground state and the excited state, with the relative population depending on the accumulated phase (see top inset). 

\begin{figure}[tbp]
\centering
\includegraphics[width=0.48\textwidth]{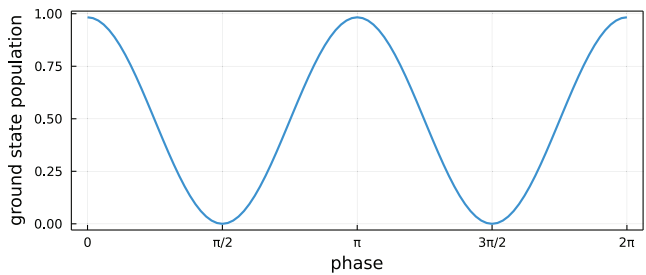}
\caption{%
Ground state population of the tractor potential at the final time as a function of the accumulated differential phase between the left and the right arm of the TAI. The phase was modeled by introducing an instantaneous phase kick to the wave function in the left potential at $t = 10~$ms $+ \, \tau/2$, see Fig.~\ref{Fig4}.
}
\label{Fig5}
\end{figure}

To demonstrate the performance of the complete scalar TAI interferometric scheme, we propagate the atomic wave packet according to Eq.~(\ref{Schroedinger}). After splitting ($T_s = 10$~ms in Fig.~\ref{Fig4}), we add an instantaneous phase kick to the wave-function component in the left tractor potential and then calculate the ground state population after recombination ($t = 20~$ms $+ \, \tau$ in Fig.~\ref{Fig4}). In Fig.~\ref{Fig5}, we plot the population of the ground state as a function of the interferometric phase. We obtain a sinusoidal function as expected for the device. 
A loss of contrast on the order of 2\% due to residual population in other levels is observed. However, the sensitivity gain will be more significant than the contrast loss as the speed-up allows to maintain the atoms separated at the maximum distance for a longer time. Moreover, the contrast loss can be further reduced by applying, for example, a penalty on the population of unwanted states or more advanced optimal control techniques~\cite{SolaAAMOP2018AsPaper,goerz2022quantum}.

\section{Discussion and Conclusions}
\label{sec:discussion}

In this work, we describe a concept and some key features of tractor atom interferometry based on 3D-confined guiding of cold atoms through controllable predetermined channels. Several implementation schemes of TAI have been discussed to provide a way to estimate AI performance. Microgravity will afford the use of extremely shallow and large super-relaxed TAI wells with a virtual absence of tunneling. Such features promise high levels of performance to address topics in the areas of fundamental physics and low-$g$ navigation. 
``Magic'' optical-tractor wells with depths in the sub-Hz regime, prepared with intensity-leveled trapping beams, are expected to afford splitting times exceeding several minutes, as well as macroscopic distances between coherently split wave-function components. These capabilities, combined with dispersion-free, 3D uninterrupted wave-function confinement shared among all TAI implementations, and with quantum correlated (entangled or spin-squeezed) atomic ensembles~\cite{PRApp2022} to move towards the Heisenberg quantum-sensing limit, form prerequisites for future transformative progress in high-precision AI.

TAI offers compactness and can be used as various inertial sensors to measure acceleration, gravity, and rotation. For a performance estimate of acceleration sensitivity, we use $T=100$ s and $z=0.1$ m and a phase resolution of $2\pi/100$ to find  $\delta a = 5 \times 10^{-13} g$. The spinor case also allows for introducing spin squeezing, which can further improve sensitivity by an extra order of magnitude, reaching $\delta a = 5 \times 10^{-14} g$. These figures surpass state-of-the-art sensitivities ($~10^{-9} g$~\cite{Menoret2018}) due to the long hold time afforded by TAI and the squeezing. For angular frequency sensitivity, we assume an area of $A=0.01$~m$^2$, a phase resolution of $2\pi/100$ and $K=10$ loops (which can be traversed within $100$ s) to find a resolution of $\delta \Omega = 2 \times 10^{-10}$ rad/s, which improves to $\delta \Omega = 2 \times 10^{-11}$ rad/s employing spin-squeezed atomic states. These estimates also compare favorably to the state of the art.

Regarding coherent quantum control, we have demonstrated its ability to speed up the splitting and recombination dynamics for the spin-less case by two orders of magnitude. Consequently, we foresee at least a similar speed-up for the spinor and Sagnac interferometers, where nonadiabaticity also plays a role. Moreover, one could incorporate more elaborate numeric approaches and functionals to maximize fidelity and increase robustness due to imperfections in the controls.

The precision of AIs enables the search for dark matter~\cite{safronovarmp2018, Badurina_2020}, the detection of gravitational waves~\cite{yu2011, Hogan2016, canuel2018, Badurina_2020, Abe_2021}, and inertial navigation~\cite{Bongs2019}. There is considerable untapped potential in further advances within these devices. Improvements in atomic clocks would significantly enhance the long-term stability of navigation and timekeeping in situations where communication and re-synchronization are limited. AI-based gravity gradiometers can reach sensitivities of several orders of magnitude beyond conventional gravity sensors, allowing them to detect dense materials in shipping containers or underground structures. We believe that AI-based sensors for navigation, fundamental physics research, and remote probing of celestial bodies via AI-based geodesy will play a crucial role in future NASA missions.

\section*{ACKNOWLEDGMENTS}
The work at the University of Michigan was supported by the NSF Grant No. PHY-2110049. MHG and SCC acknowledge support by the DEVCOM Army Research Laboratory under Cooperative Agreement Number W911NF-16-2-0147 and W911NF-21-2-0037, respectively. VSM is grateful for support by a Laboratory University Collaboration Initiative (LUCI) grant from OUSD.

\bibliography{references}

\begin{thebibliography}{72}%
\makeatletter
\providecommand \@ifxundefined [1]{%
 \@ifx{#1\undefined}
}%
\providecommand \@ifnum [1]{%
 \ifnum #1\expandafter \@firstoftwo
 \else \expandafter \@secondoftwo
 \fi
}%
\providecommand \@ifx [1]{%
 \ifx #1\expandafter \@firstoftwo
 \else \expandafter \@secondoftwo
 \fi
}%
\providecommand \natexlab [1]{#1}%
\providecommand \enquote  [1]{``#1''}%
\providecommand \bibnamefont  [1]{#1}%
\providecommand \bibfnamefont [1]{#1}%
\providecommand \citenamefont [1]{#1}%
\providecommand \href@noop [0]{\@secondoftwo}%
\providecommand \href [0]{\begingroup \@sanitize@url \@href}%
\providecommand \@href[1]{\@@startlink{#1}\@@href}%
\providecommand \@@href[1]{\endgroup#1\@@endlink}%
\providecommand \@sanitize@url [0]{\catcode `\\12\catcode `\$12\catcode
  `\&12\catcode `\#12\catcode `\^12\catcode `\_12\catcode `\%12\relax}%
\providecommand \@@startlink[1]{}%
\providecommand \@@endlink[0]{}%
\providecommand \url  [0]{\begingroup\@sanitize@url \@url }%
\providecommand \@url [1]{\endgroup\@href {#1}{\urlprefix }}%
\providecommand \urlprefix  [0]{URL }%
\providecommand \Eprint [0]{\href }%
\providecommand \doibase [0]{http://dx.doi.org/}%
\providecommand \selectlanguage [0]{\@gobble}%
\providecommand \bibinfo  [0]{\@secondoftwo}%
\providecommand \bibfield  [0]{\@secondoftwo}%
\providecommand \translation [1]{[#1]}%
\providecommand \BibitemOpen [0]{}%
\providecommand \bibitemStop [0]{}%
\providecommand \bibitemNoStop [0]{.\EOS\space}%
\providecommand \EOS [0]{\spacefactor3000\relax}%
\providecommand \BibitemShut  [1]{\csname bibitem#1\endcsname}%
\let\auto@bib@innerbib\@empty
%</preamble>
\bibitem [{\citenamefont {Carnal}\ and\ \citenamefont
  {Mlynek}(1991)}]{Carnal1991}%
  \BibitemOpen
  \bibfield  {author} {\bibinfo {author} {\bibfnamefont {O.}~\bibnamefont
  {Carnal}}\ and\ \bibinfo {author} {\bibfnamefont {J.}~\bibnamefont
  {Mlynek}},\ }\bibfield  {title} {\enquote {\bibinfo {title} {Young's
  double-slit experiment with atoms: A simple atom interferometer},}\ }\href
  {\doibase 10.1103/PhysRevLett.66.2689} {\bibfield  {journal} {\bibinfo
  {journal} {Phys. Rev. Lett.}\ }\textbf {\bibinfo {volume} {66}},\ \bibinfo
  {pages} {2689--2692} (\bibinfo {year} {1991})}\BibitemShut {NoStop}%
\bibitem [{\citenamefont {Keith}\ \emph {et~al.}(1991)\citenamefont {Keith},
  \citenamefont {Ekstrom}, \citenamefont {Turchette},\ and\ \citenamefont
  {Pritchard}}]{Keith1991}%
  \BibitemOpen
  \bibfield  {author} {\bibinfo {author} {\bibfnamefont {D.~W.}\ \bibnamefont
  {Keith}}, \bibinfo {author} {\bibfnamefont {C.~R.}\ \bibnamefont {Ekstrom}},
  \bibinfo {author} {\bibfnamefont {Q.~A.}\ \bibnamefont {Turchette}}, \ and\
  \bibinfo {author} {\bibfnamefont {D.~E.}\ \bibnamefont {Pritchard}},\
  }\bibfield  {title} {\enquote {\bibinfo {title} {An interferometer for
  atoms},}\ }\href {\doibase 10.1103/PhysRevLett.66.2693} {\bibfield  {journal}
  {\bibinfo  {journal} {Phys. Rev. Lett.}\ }\textbf {\bibinfo {volume} {66}},\
  \bibinfo {pages} {2693--2696} (\bibinfo {year} {1991})}\BibitemShut {NoStop}%
\bibitem [{\citenamefont {Riehle}\ \emph {et~al.}(1991)\citenamefont {Riehle},
  \citenamefont {Kisters}, \citenamefont {Witte}, \citenamefont {Helmcke},\
  and\ \citenamefont {Bord\'e}}]{Riehle1991}%
  \BibitemOpen
  \bibfield  {author} {\bibinfo {author} {\bibfnamefont {F.}~\bibnamefont
  {Riehle}}, \bibinfo {author} {\bibfnamefont {Th.}\ \bibnamefont {Kisters}},
  \bibinfo {author} {\bibfnamefont {A.}~\bibnamefont {Witte}}, \bibinfo
  {author} {\bibfnamefont {J.}~\bibnamefont {Helmcke}}, \ and\ \bibinfo
  {author} {\bibfnamefont {Ch.~J.}\ \bibnamefont {Bord\'e}},\ }\bibfield
  {title} {\enquote {\bibinfo {title} {Optical ramsey spectroscopy in a
  rotating frame: {S}agnac effect in a matter-wave interferometer},}\ }\href
  {\doibase 10.1103/PhysRevLett.67.177} {\bibfield  {journal} {\bibinfo
  {journal} {Phys. Rev. Lett.}\ }\textbf {\bibinfo {volume} {67}},\ \bibinfo
  {pages} {177--180} (\bibinfo {year} {1991})}\BibitemShut {NoStop}%
\bibitem [{\citenamefont {Kasevich}\ and\ \citenamefont
  {Chu}(1991)}]{Kasevich1991}%
  \BibitemOpen
  \bibfield  {author} {\bibinfo {author} {\bibfnamefont {M.}~\bibnamefont
  {Kasevich}}\ and\ \bibinfo {author} {\bibfnamefont {S.}~\bibnamefont {Chu}},\
  }\bibfield  {title} {\enquote {\bibinfo {title} {Atomic interferometry using
  stimulated {R}aman transitions},}\ }\href {\doibase
  10.1103/PhysRevLett.67.181} {\bibfield  {journal} {\bibinfo  {journal} {Phys.
  Rev. Lett.}\ }\textbf {\bibinfo {volume} {67}},\ \bibinfo {pages} {181--184}
  (\bibinfo {year} {1991})}\BibitemShut {NoStop}%
\bibitem [{\citenamefont {Cronin}\ \emph {et~al.}(2009)\citenamefont {Cronin},
  \citenamefont {Schmiedmayer},\ and\ \citenamefont {Pritchard}}]{Cronin2009}%
  \BibitemOpen
  \bibfield  {author} {\bibinfo {author} {\bibfnamefont {A.~D.}\ \bibnamefont
  {Cronin}}, \bibinfo {author} {\bibfnamefont {J.}~\bibnamefont
  {Schmiedmayer}}, \ and\ \bibinfo {author} {\bibfnamefont {D.~E.}\
  \bibnamefont {Pritchard}},\ }\bibfield  {title} {\enquote {\bibinfo {title}
  {Optics and interferometry with atoms and molecules},}\ }\href {\doibase
  10.1103/RevModPhys.81.1051} {\bibfield  {journal} {\bibinfo  {journal} {Rev.
  Mod. Phys.}\ }\textbf {\bibinfo {volume} {81}},\ \bibinfo {pages}
  {1051--1129} (\bibinfo {year} {2009})}\BibitemShut {NoStop}%
\bibitem [{\citenamefont {Berman}(1997)}]{Berman1997}%
  \BibitemOpen
  \bibfield  {author} {\bibinfo {author} {\bibfnamefont {P.}~\bibnamefont
  {Berman}},\ }\href@noop {} {\emph {\bibinfo {title} {Atom interferometry}}}\
  (\bibinfo  {publisher} {Academic Press},\ \bibinfo {address} {San Diego},\
  \bibinfo {year} {1997})\BibitemShut {NoStop}%
\bibitem [{\citenamefont {Abend}\ \emph {et~al.}(2020)\citenamefont {Abend},
  \citenamefont {Gersemann}, \citenamefont {Schubert}, \citenamefont
  {Schlippert}, \citenamefont {Rasel}, \citenamefont {Zimmermann},
  \citenamefont {Efremov}, \citenamefont {Roura}, \citenamefont {Narducci},\
  and\ \citenamefont {Schleich}}]{Abend2020}%
  \BibitemOpen
  \bibfield  {author} {\bibinfo {author} {\bibfnamefont {S.}~\bibnamefont
  {Abend}}, \bibinfo {author} {\bibfnamefont {M.}~\bibnamefont {Gersemann}},
  \bibinfo {author} {\bibfnamefont {D.}~\bibnamefont {Schubert}}, \bibinfo
  {author} {\bibfnamefont {C.}~\bibnamefont {Schlippert}}, \bibinfo {author}
  {\bibfnamefont {E.~M.}\ \bibnamefont {Rasel}}, \bibinfo {author}
  {\bibfnamefont {M.}~\bibnamefont {Zimmermann}}, \bibinfo {author}
  {\bibfnamefont {M.~A.}\ \bibnamefont {Efremov}}, \bibinfo {author}
  {\bibfnamefont {A.}~\bibnamefont {Roura}}, \bibinfo {author} {\bibfnamefont
  {F.~A.}\ \bibnamefont {Narducci}}, \ and\ \bibinfo {author} {\bibfnamefont
  {W.~P.}\ \bibnamefont {Schleich}},\ }\bibfield  {title} {\enquote {\bibinfo
  {title} {Atom interferometry and its applications},}\ }in\ \href
  {https://ebooks.iospress.nl/pdf/doi/10.3254/978-1-61499-937-9-345} {\emph
  {\bibinfo {booktitle} {Proceedings of the International School of Physics
  Enrico Fermi}}},\ \bibinfo {editor} {edited by\ \bibinfo {editor}
  {\bibfnamefont {W.~P.}\ \bibnamefont {Schleich}}, \bibinfo {editor}
  {\bibfnamefont {E.~M.}\ \bibnamefont {Rasel}}, \ and\ \bibinfo {editor}
  {\bibfnamefont {S.}~\bibnamefont {Wolk}}}\ (\bibinfo  {publisher} {IOS
  Press},\ \bibinfo {address} {Amsterdam},\ \bibinfo {year} {2020})\BibitemShut
  {NoStop}%
\bibitem [{\citenamefont {Tarallo}\ \emph {et~al.}(2014)\citenamefont
  {Tarallo}, \citenamefont {Mazzoni}, \citenamefont {Poli}, \citenamefont
  {Sutyrin}, \citenamefont {Zhang},\ and\ \citenamefont {Tino}}]{Tarallo2014}%
  \BibitemOpen
  \bibfield  {author} {\bibinfo {author} {\bibfnamefont {M.~G.}\ \bibnamefont
  {Tarallo}}, \bibinfo {author} {\bibfnamefont {T.}~\bibnamefont {Mazzoni}},
  \bibinfo {author} {\bibfnamefont {N.}~\bibnamefont {Poli}}, \bibinfo {author}
  {\bibfnamefont {D.~V.}\ \bibnamefont {Sutyrin}}, \bibinfo {author}
  {\bibfnamefont {X.}~\bibnamefont {Zhang}}, \ and\ \bibinfo {author}
  {\bibfnamefont {G.~M.}\ \bibnamefont {Tino}},\ }\bibfield  {title} {\enquote
  {\bibinfo {title} {Test of {E}instein equivalence principle for 0-spin and
  half-integer-spin atoms: Search for spin-gravity coupling effects},}\ }\href
  {\doibase 10.1103/PhysRevLett.113.023005} {\bibfield  {journal} {\bibinfo
  {journal} {Phys. Rev. Lett.}\ }\textbf {\bibinfo {volume} {113}},\ \bibinfo
  {pages} {023005} (\bibinfo {year} {2014})}\BibitemShut {NoStop}%
\bibitem [{\citenamefont {Schlippert}\ \emph {et~al.}(2014)\citenamefont
  {Schlippert}, \citenamefont {Hartwig}, \citenamefont {Albers}, \citenamefont
  {Richardson}, \citenamefont {Schubert}, \citenamefont {Roura}, \citenamefont
  {Schleich}, \citenamefont {Ertmer},\ and\ \citenamefont
  {Rasel}}]{Schlippert2014}%
  \BibitemOpen
  \bibfield  {author} {\bibinfo {author} {\bibfnamefont {D.}~\bibnamefont
  {Schlippert}}, \bibinfo {author} {\bibfnamefont {J.}~\bibnamefont {Hartwig}},
  \bibinfo {author} {\bibfnamefont {H.}~\bibnamefont {Albers}}, \bibinfo
  {author} {\bibfnamefont {L.~L.}\ \bibnamefont {Richardson}}, \bibinfo
  {author} {\bibfnamefont {C.}~\bibnamefont {Schubert}}, \bibinfo {author}
  {\bibfnamefont {A.}~\bibnamefont {Roura}}, \bibinfo {author} {\bibfnamefont
  {W.~P.}\ \bibnamefont {Schleich}}, \bibinfo {author} {\bibfnamefont
  {W.}~\bibnamefont {Ertmer}}, \ and\ \bibinfo {author} {\bibfnamefont {E.~M.}\
  \bibnamefont {Rasel}},\ }\bibfield  {title} {\enquote {\bibinfo {title}
  {Quantum test of the universality of free fall},}\ }\href {\doibase
  10.1103/PhysRevLett.112.203002} {\bibfield  {journal} {\bibinfo  {journal}
  {Phys. Rev. Lett.}\ }\textbf {\bibinfo {volume} {112}},\ \bibinfo {pages}
  {203002} (\bibinfo {year} {2014})}\BibitemShut {NoStop}%
\bibitem [{\citenamefont {Kovachy}\ \emph {et~al.}(2015)\citenamefont
  {Kovachy}, \citenamefont {Asenbaum}, \citenamefont {Overstreet},
  \citenamefont {Donnelly}, \citenamefont {Dickerson}, \citenamefont
  {Sugarbaker}, \citenamefont {Hogan},\ and\ \citenamefont
  {Kasevich}}]{Kovachy2015}%
  \BibitemOpen
  \bibfield  {author} {\bibinfo {author} {\bibfnamefont {T.}~\bibnamefont
  {Kovachy}}, \bibinfo {author} {\bibfnamefont {P.}~\bibnamefont {Asenbaum}},
  \bibinfo {author} {\bibfnamefont {C.}~\bibnamefont {Overstreet}}, \bibinfo
  {author} {\bibfnamefont {C.~A.}\ \bibnamefont {Donnelly}}, \bibinfo {author}
  {\bibfnamefont {S.~M.}\ \bibnamefont {Dickerson}}, \bibinfo {author}
  {\bibfnamefont {A.}~\bibnamefont {Sugarbaker}}, \bibinfo {author}
  {\bibfnamefont {J.~M.}\ \bibnamefont {Hogan}}, \ and\ \bibinfo {author}
  {\bibfnamefont {M.~A.}\ \bibnamefont {Kasevich}},\ }\bibfield  {title}
  {\enquote {\bibinfo {title} {Quantum superposition at the half-metre
  scale},}\ }\href {\doibase 10.1038/nature16155} {\bibfield  {journal}
  {\bibinfo  {journal} {Nature}\ }\textbf {\bibinfo {volume} {528}},\ \bibinfo
  {pages} {530--533} (\bibinfo {year} {2015})}\BibitemShut {NoStop}%
\bibitem [{\citenamefont {Jaffe}\ \emph {et~al.}(2017)\citenamefont {Jaffe},
  \citenamefont {Haslinger}, \citenamefont {Xu}, \citenamefont {Hamilton},
  \citenamefont {Upadhye}, \citenamefont {Elder}, \citenamefont {Khoury},\ and\
  \citenamefont {M{\"u}ller}}]{Jaffe2017}%
  \BibitemOpen
  \bibfield  {author} {\bibinfo {author} {\bibfnamefont {M.}~\bibnamefont
  {Jaffe}}, \bibinfo {author} {\bibfnamefont {P.}~\bibnamefont {Haslinger}},
  \bibinfo {author} {\bibfnamefont {V.}~\bibnamefont {Xu}}, \bibinfo {author}
  {\bibfnamefont {P.}~\bibnamefont {Hamilton}}, \bibinfo {author}
  {\bibfnamefont {A.}~\bibnamefont {Upadhye}}, \bibinfo {author} {\bibfnamefont
  {B.}~\bibnamefont {Elder}}, \bibinfo {author} {\bibfnamefont
  {J.}~\bibnamefont {Khoury}}, \ and\ \bibinfo {author} {\bibfnamefont
  {H.}~\bibnamefont {M{\"u}ller}},\ }\bibfield  {title} {\enquote {\bibinfo
  {title} {Testing sub-gravitational forces on atoms from a miniature in-vacuum
  source mass},}\ }\href {\doibase 10.1038/nphys4189} {\bibfield  {journal}
  {\bibinfo  {journal} {Nat. Phys.}\ }\textbf {\bibinfo {volume} {13}},\
  \bibinfo {pages} {938--942} (\bibinfo {year} {2017})}\BibitemShut {NoStop}%
\bibitem [{\citenamefont {Rosi}\ \emph {et~al.}(2017)\citenamefont {Rosi},
  \citenamefont {D'Amico}, \citenamefont {Cacciapuoti}, \citenamefont
  {Sorrentino}, \citenamefont {Prevedelli}, \citenamefont {Zych}, \citenamefont
  {Brukner},\ and\ \citenamefont {Tino}}]{Rosi2017}%
  \BibitemOpen
  \bibfield  {author} {\bibinfo {author} {\bibfnamefont {G.}~\bibnamefont
  {Rosi}}, \bibinfo {author} {\bibfnamefont {G.}~\bibnamefont {D'Amico}},
  \bibinfo {author} {\bibfnamefont {L.}~\bibnamefont {Cacciapuoti}}, \bibinfo
  {author} {\bibfnamefont {F.}~\bibnamefont {Sorrentino}}, \bibinfo {author}
  {\bibfnamefont {M.}~\bibnamefont {Prevedelli}}, \bibinfo {author}
  {\bibfnamefont {M.}~\bibnamefont {Zych}}, \bibinfo {author} {\bibfnamefont
  {{\v{C}}}~\bibnamefont {Brukner}}, \ and\ \bibinfo {author} {\bibfnamefont
  {G.~M.}\ \bibnamefont {Tino}},\ }\bibfield  {title} {\enquote {\bibinfo
  {title} {Quantum test of the equivalence principle for atoms in coherent
  superposition of internal energy states},}\ }\href {\doibase
  10.1038/ncomms15529} {\bibfield  {journal} {\bibinfo  {journal} {Nature
  Comm.}\ }\textbf {\bibinfo {volume} {8}},\ \bibinfo {pages} {15529} (\bibinfo
  {year} {2017})}\BibitemShut {NoStop}%
\bibitem [{\citenamefont {Fixler}\ \emph {et~al.}(2007)\citenamefont {Fixler},
  \citenamefont {Foster}, \citenamefont {McGuirk},\ and\ \citenamefont
  {Kasevich}}]{Fixler2007}%
  \BibitemOpen
  \bibfield  {author} {\bibinfo {author} {\bibfnamefont {J.~B.}\ \bibnamefont
  {Fixler}}, \bibinfo {author} {\bibfnamefont {G.~T.}\ \bibnamefont {Foster}},
  \bibinfo {author} {\bibfnamefont {J.~M.}\ \bibnamefont {McGuirk}}, \ and\
  \bibinfo {author} {\bibfnamefont {M.~A.}\ \bibnamefont {Kasevich}},\
  }\bibfield  {title} {\enquote {\bibinfo {title} {Atom interferometer
  measurement of the {N}ewtonian constant of gravity},}\ }\href {\doibase
  10.1126/science.1135459} {\bibfield  {journal} {\bibinfo  {journal}
  {Science}\ }\textbf {\bibinfo {volume} {315}},\ \bibinfo {pages} {74--77}
  (\bibinfo {year} {2007})}\BibitemShut {NoStop}%
\bibitem [{\citenamefont {M\'{e}noret}\ \emph {et~al.}(2018)\citenamefont
  {M\'{e}noret}, \citenamefont {Vermeulen}, \citenamefont {Le~Moigne},
  \citenamefont {Bonvalot}, \citenamefont {Bouyer}, \citenamefont {Landragin},\
  and\ \citenamefont {Desruelle}}]{Menoret2018}%
  \BibitemOpen
  \bibfield  {author} {\bibinfo {author} {\bibfnamefont {V.}~\bibnamefont
  {M\'{e}noret}}, \bibinfo {author} {\bibfnamefont {P.}~\bibnamefont
  {Vermeulen}}, \bibinfo {author} {\bibfnamefont {N.}~\bibnamefont
  {Le~Moigne}}, \bibinfo {author} {\bibfnamefont {S.}~\bibnamefont {Bonvalot}},
  \bibinfo {author} {\bibfnamefont {P.}~\bibnamefont {Bouyer}}, \bibinfo
  {author} {\bibfnamefont {A.}~\bibnamefont {Landragin}}, \ and\ \bibinfo
  {author} {\bibfnamefont {B.}~\bibnamefont {Desruelle}},\ }\bibfield  {title}
  {\enquote {\bibinfo {title} {Gravity measurements below 10$^{-9}$g with a
  transportable absolute quantum gravimeter},}\ }\href {\doibase
  10.1038/s41598-018-30608-1} {\bibfield  {journal} {\bibinfo  {journal} {Sci.
  Rep.}\ }\textbf {\bibinfo {volume} {8}},\ \bibinfo {pages} {12300} (\bibinfo
  {year} {2018})}\BibitemShut {NoStop}%
\bibitem [{\citenamefont {Xu}\ \emph {et~al.}(2019)\citenamefont {Xu},
  \citenamefont {Jaffe}, \citenamefont {Panda}, \citenamefont {Kristensen},
  \citenamefont {Clark},\ and\ \citenamefont {M\''üller}}]{Xu2019}%
  \BibitemOpen
  \bibfield  {author} {\bibinfo {author} {\bibfnamefont {V.}~\bibnamefont
  {Xu}}, \bibinfo {author} {\bibfnamefont {M.}~\bibnamefont {Jaffe}}, \bibinfo
  {author} {\bibfnamefont {C.~D.}\ \bibnamefont {Panda}}, \bibinfo {author}
  {\bibfnamefont {S.~L.}\ \bibnamefont {Kristensen}}, \bibinfo {author}
  {\bibfnamefont {L.~W.}\ \bibnamefont {Clark}}, \ and\ \bibinfo {author}
  {\bibfnamefont {H.}~\bibnamefont {M\''üller}},\ }\bibfield  {title}
  {\enquote {\bibinfo {title} {Probing gravity by holding atoms for 20
  seconds},}\ }\href {\doibase 10.1126/science.aay6428} {\bibfield  {journal}
  {\bibinfo  {journal} {Science}\ }\textbf {\bibinfo {volume} {366}},\ \bibinfo
  {pages} {745--749} (\bibinfo {year} {2019})}\BibitemShut {NoStop}%
\bibitem [{\citenamefont {Hogan}\ and\ \citenamefont
  {Kasevich}(2016)}]{Hogan2016}%
  \BibitemOpen
  \bibfield  {author} {\bibinfo {author} {\bibfnamefont {J.~M.}\ \bibnamefont
  {Hogan}}\ and\ \bibinfo {author} {\bibfnamefont {M.~A.}\ \bibnamefont
  {Kasevich}},\ }\bibfield  {title} {\enquote {\bibinfo {title}
  {Atom-interferometric gravitational-wave detection using heterodyne laser
  links},}\ }\href {\doibase 10.1103/PhysRevA.94.033632} {\bibfield  {journal}
  {\bibinfo  {journal} {Phys. Rev. A}\ }\textbf {\bibinfo {volume} {94}},\
  \bibinfo {pages} {033632} (\bibinfo {year} {2016})}\BibitemShut {NoStop}%
\bibitem [{\citenamefont {Hanneke}\ \emph {et~al.}(2008)\citenamefont
  {Hanneke}, \citenamefont {Fogwell},\ and\ \citenamefont
  {Gabrielse}}]{Hanneke2008}%
  \BibitemOpen
  \bibfield  {author} {\bibinfo {author} {\bibfnamefont {D.}~\bibnamefont
  {Hanneke}}, \bibinfo {author} {\bibfnamefont {S.}~\bibnamefont {Fogwell}}, \
  and\ \bibinfo {author} {\bibfnamefont {G.}~\bibnamefont {Gabrielse}},\
  }\bibfield  {title} {\enquote {\bibinfo {title} {New measurement of the
  electron magnetic moment and the fine structure constant},}\ }\href {\doibase
  10.1103/PhysRevLett.100.120801} {\bibfield  {journal} {\bibinfo  {journal}
  {Phys. Rev. Lett.}\ }\textbf {\bibinfo {volume} {100}},\ \bibinfo {pages}
  {120801} (\bibinfo {year} {2008})}\BibitemShut {NoStop}%
\bibitem [{\citenamefont {Parker}\ \emph {et~al.}(2018)\citenamefont {Parker},
  \citenamefont {Yu}, \citenamefont {Zhong}, \citenamefont {Estey},\ and\
  \citenamefont {Müller}}]{Parker2018}%
  \BibitemOpen
  \bibfield  {author} {\bibinfo {author} {\bibfnamefont {R.~H.}\ \bibnamefont
  {Parker}}, \bibinfo {author} {\bibfnamefont {C.}~\bibnamefont {Yu}}, \bibinfo
  {author} {\bibfnamefont {W.}~\bibnamefont {Zhong}}, \bibinfo {author}
  {\bibfnamefont {B.}~\bibnamefont {Estey}}, \ and\ \bibinfo {author}
  {\bibfnamefont {H.}~\bibnamefont {Müller}},\ }\bibfield  {title} {\enquote
  {\bibinfo {title} {Measurement of the fine-structure constant as a test of
  the {S}tandard {M}odel},}\ }\href {\doibase 10.1126/science.aap7706}
  {\bibfield  {journal} {\bibinfo  {journal} {Science}\ }\textbf {\bibinfo
  {volume} {360}},\ \bibinfo {pages} {191--195} (\bibinfo {year}
  {2018})}\BibitemShut {NoStop}%
\bibitem [{\citenamefont {Morel}\ \emph {et~al.}(2020)\citenamefont {Morel},
  \citenamefont {Yao}, \citenamefont {Clad{\'e}},\ and\ \citenamefont
  {Guellati-Kh{\'e}lifa}}]{Morel2020}%
  \BibitemOpen
  \bibfield  {author} {\bibinfo {author} {\bibfnamefont {L.}~\bibnamefont
  {Morel}}, \bibinfo {author} {\bibfnamefont {Z.}~\bibnamefont {Yao}}, \bibinfo
  {author} {\bibfnamefont {P.}~\bibnamefont {Clad{\'e}}}, \ and\ \bibinfo
  {author} {\bibfnamefont {S.}~\bibnamefont {Guellati-Kh{\'e}lifa}},\
  }\bibfield  {title} {\enquote {\bibinfo {title} {Determination of the
  fine-structure constant with an accuracy of 81 parts per trillion},}\ }\href
  {\doibase 10.1038/s41586-020-2964-7} {\bibfield  {journal} {\bibinfo
  {journal} {Nature}\ }\textbf {\bibinfo {volume} {588}},\ \bibinfo {pages}
  {61--65} (\bibinfo {year} {2020})}\BibitemShut {NoStop}%
\bibitem [{\citenamefont {Barrett}\ \emph {et~al.}(2019)\citenamefont
  {Barrett}, \citenamefont {Cheiney}, \citenamefont {Battelier}, \citenamefont
  {Napolitano},\ and\ \citenamefont {Bouyer}}]{Barrett2019}%
  \BibitemOpen
  \bibfield  {author} {\bibinfo {author} {\bibfnamefont {B.}~\bibnamefont
  {Barrett}}, \bibinfo {author} {\bibfnamefont {P.}~\bibnamefont {Cheiney}},
  \bibinfo {author} {\bibfnamefont {B.}~\bibnamefont {Battelier}}, \bibinfo
  {author} {\bibfnamefont {F.}~\bibnamefont {Napolitano}}, \ and\ \bibinfo
  {author} {\bibfnamefont {P.}~\bibnamefont {Bouyer}},\ }\bibfield  {title}
  {\enquote {\bibinfo {title} {Multidimensional atom optics and
  interferometry},}\ }\href {\doibase 10.1103/PhysRevLett.122.043604}
  {\bibfield  {journal} {\bibinfo  {journal} {Phys. Rev. Lett.}\ }\textbf
  {\bibinfo {volume} {122}},\ \bibinfo {pages} {043604} (\bibinfo {year}
  {2019})}\BibitemShut {NoStop}%
\bibitem [{\citenamefont {Bongs}\ \emph {et~al.}(2019)\citenamefont {Bongs},
  \citenamefont {Holynski}, \citenamefont {Vovrosh}, \citenamefont {Bouyer},
  \citenamefont {Condon}, \citenamefont {Rasel}, \citenamefont {Schubert},
  \citenamefont {Schleich},\ and\ \citenamefont {Roura}}]{Bongs2019}%
  \BibitemOpen
  \bibfield  {author} {\bibinfo {author} {\bibfnamefont {K.}~\bibnamefont
  {Bongs}}, \bibinfo {author} {\bibfnamefont {M.}~\bibnamefont {Holynski}},
  \bibinfo {author} {\bibfnamefont {J.}~\bibnamefont {Vovrosh}}, \bibinfo
  {author} {\bibfnamefont {P.}~\bibnamefont {Bouyer}}, \bibinfo {author}
  {\bibfnamefont {G.}~\bibnamefont {Condon}}, \bibinfo {author} {\bibfnamefont
  {E.}~\bibnamefont {Rasel}}, \bibinfo {author} {\bibfnamefont
  {C.}~\bibnamefont {Schubert}}, \bibinfo {author} {\bibfnamefont {W.~P.}\
  \bibnamefont {Schleich}}, \ and\ \bibinfo {author} {\bibfnamefont
  {A.}~\bibnamefont {Roura}},\ }\bibfield  {title} {\enquote {\bibinfo {title}
  {Taking atom interferometric quantum sensors from the laboratory to
  real-world applications},}\ }\href {\doibase 10.1038/s42254-019-0117-4}
  {\bibfield  {journal} {\bibinfo  {journal} {Nat. Rev. Phys.}\ }\textbf
  {\bibinfo {volume} {1}},\ \bibinfo {pages} {731--739} (\bibinfo {year}
  {2019})}\BibitemShut {NoStop}%
\bibitem [{\citenamefont {Garrido~Alzar}(2019)}]{Alzar2019}%
  \BibitemOpen
  \bibfield  {author} {\bibinfo {author} {\bibfnamefont {C.~L.}\ \bibnamefont
  {Garrido~Alzar}},\ }\bibfield  {title} {\enquote {\bibinfo {title} {Compact
  chip-scale guided cold atom gyrometers for inertial navigation: Enabling
  technologies and design study},}\ }\href {\doibase 10.1116/1.5120348}
  {\bibfield  {journal} {\bibinfo  {journal} {AVS Quantum Science}\ }\textbf
  {\bibinfo {volume} {1}},\ \bibinfo {pages} {014702} (\bibinfo {year}
  {2019})}\BibitemShut {NoStop}%
\bibitem [{\citenamefont {Bidel}\ \emph {et~al.}(2020)\citenamefont {Bidel},
  \citenamefont {Zahzam}, \citenamefont {Bresson}, \citenamefont {Blanchard},
  \citenamefont {Cadoret}, \citenamefont {Olesen},\ and\ \citenamefont
  {Forsberg}}]{Bidel2020}%
  \BibitemOpen
  \bibfield  {author} {\bibinfo {author} {\bibfnamefont {Y.}~\bibnamefont
  {Bidel}}, \bibinfo {author} {\bibfnamefont {N.}~\bibnamefont {Zahzam}},
  \bibinfo {author} {\bibfnamefont {A.}~\bibnamefont {Bresson}}, \bibinfo
  {author} {\bibfnamefont {C.}~\bibnamefont {Blanchard}}, \bibinfo {author}
  {\bibfnamefont {M.}~\bibnamefont {Cadoret}}, \bibinfo {author} {\bibfnamefont
  {A.~V.}\ \bibnamefont {Olesen}}, \ and\ \bibinfo {author} {\bibfnamefont
  {R.}~\bibnamefont {Forsberg}},\ }\bibfield  {title} {\enquote {\bibinfo
  {title} {Absolute airborne gravimetry with a cold atom sensor},}\ }\href
  {\doibase 10.1007/s00190-020-01350-2} {\bibfield  {journal} {\bibinfo
  {journal} {J. Geod.}\ }\textbf {\bibinfo {volume} {94}} (\bibinfo {year}
  {2020}),\ 10.1007/s00190-020-01350-2}\BibitemShut {NoStop}%
\bibitem [{\citenamefont {et~al.}(2019)}]{Tino2019}%
  \BibitemOpen
  \bibfield  {author} {\bibinfo {author} {\bibfnamefont {G.~M.~Tino}\
  \bibnamefont {et~al.}},\ }\bibfield  {title} {\enquote {\bibinfo {title}
  {{SAGE}: A proposal for a space atomic gravity explorer},}\ }\href {\doibase
  10.1140/epjd/e2019-100324-6} {\bibfield  {journal} {\bibinfo  {journal} {Eur.
  Phys. J. D}\ }\textbf {\bibinfo {volume} {73}},\ \bibinfo {pages} {228}
  (\bibinfo {year} {2019})}\BibitemShut {NoStop}%
\bibitem [{\citenamefont {Asenbaum}\ \emph {et~al.}(2017)\citenamefont
  {Asenbaum}, \citenamefont {Overstreet}, \citenamefont {Kovachy},
  \citenamefont {Brown}, \citenamefont {Hogan},\ and\ \citenamefont
  {Kasevich}}]{Asenbaum2017}%
  \BibitemOpen
  \bibfield  {author} {\bibinfo {author} {\bibfnamefont {P.}~\bibnamefont
  {Asenbaum}}, \bibinfo {author} {\bibfnamefont {C.}~\bibnamefont
  {Overstreet}}, \bibinfo {author} {\bibfnamefont {T.}~\bibnamefont {Kovachy}},
  \bibinfo {author} {\bibfnamefont {D.~D.}\ \bibnamefont {Brown}}, \bibinfo
  {author} {\bibfnamefont {J.~M.}\ \bibnamefont {Hogan}}, \ and\ \bibinfo
  {author} {\bibfnamefont {M.~A.}\ \bibnamefont {Kasevich}},\ }\bibfield
  {title} {\enquote {\bibinfo {title} {Phase shift in an atom interferometer
  due to spacetime curvature across its wave function},}\ }\href {\doibase
  10.1103/PhysRevLett.118.183602} {\bibfield  {journal} {\bibinfo  {journal}
  {Phys. Rev. Lett.}\ }\textbf {\bibinfo {volume} {118}},\ \bibinfo {pages}
  {183602} (\bibinfo {year} {2017})}\BibitemShut {NoStop}%
\bibitem [{\citenamefont {Rosi}\ \emph {et~al.}(2014)\citenamefont {Rosi},
  \citenamefont {Sorrentino}, \citenamefont {Cacciapuoti}, \citenamefont
  {Prevedelli},\ and\ \citenamefont {Tino}}]{Rosi2014}%
  \BibitemOpen
  \bibfield  {author} {\bibinfo {author} {\bibfnamefont {G.}~\bibnamefont
  {Rosi}}, \bibinfo {author} {\bibfnamefont {F.}~\bibnamefont {Sorrentino}},
  \bibinfo {author} {\bibfnamefont {L.}~\bibnamefont {Cacciapuoti}}, \bibinfo
  {author} {\bibfnamefont {M.}~\bibnamefont {Prevedelli}}, \ and\ \bibinfo
  {author} {\bibfnamefont {G.~M.}\ \bibnamefont {Tino}},\ }\bibfield  {title}
  {\enquote {\bibinfo {title} {Precision measurement of the {N}ewtonian
  gravitational constant using cold atoms},}\ }\href {\doibase
  10.1038/nature13433} {\bibfield  {journal} {\bibinfo  {journal} {Nature}\
  }\textbf {\bibinfo {volume} {510}},\ \bibinfo {pages} {518--521} (\bibinfo
  {year} {2014})}\BibitemShut {NoStop}%
\bibitem [{\citenamefont {Hamilton}\ \emph {et~al.}(2015)\citenamefont
  {Hamilton}, \citenamefont {Jaffe}, \citenamefont {Brown}, \citenamefont
  {Maisenbacher}, \citenamefont {Estey},\ and\ \citenamefont
  {M\"uller}}]{Hamilton2015}%
  \BibitemOpen
  \bibfield  {author} {\bibinfo {author} {\bibfnamefont {P.}~\bibnamefont
  {Hamilton}}, \bibinfo {author} {\bibfnamefont {M.}~\bibnamefont {Jaffe}},
  \bibinfo {author} {\bibfnamefont {J.~M.}\ \bibnamefont {Brown}}, \bibinfo
  {author} {\bibfnamefont {L.}~\bibnamefont {Maisenbacher}}, \bibinfo {author}
  {\bibfnamefont {B.}~\bibnamefont {Estey}}, \ and\ \bibinfo {author}
  {\bibfnamefont {H.}~\bibnamefont {M\"uller}},\ }\bibfield  {title} {\enquote
  {\bibinfo {title} {Atom interferometry in an optical cavity},}\ }\href
  {\doibase 10.1103/PhysRevLett.114.100405} {\bibfield  {journal} {\bibinfo
  {journal} {Phys. Rev. Lett.}\ }\textbf {\bibinfo {volume} {114}},\ \bibinfo
  {pages} {100405} (\bibinfo {year} {2015})}\BibitemShut {NoStop}%
\bibitem [{\citenamefont {Dickerson}\ \emph {et~al.}(2013)\citenamefont
  {Dickerson}, \citenamefont {Hogan}, \citenamefont {Sugarbaker}, \citenamefont
  {Johnson},\ and\ \citenamefont {Kasevich}}]{Dickerson2013}%
  \BibitemOpen
  \bibfield  {author} {\bibinfo {author} {\bibfnamefont {S.~M.}\ \bibnamefont
  {Dickerson}}, \bibinfo {author} {\bibfnamefont {J.~M.}\ \bibnamefont
  {Hogan}}, \bibinfo {author} {\bibfnamefont {A.}~\bibnamefont {Sugarbaker}},
  \bibinfo {author} {\bibfnamefont {D.~M.~S.}\ \bibnamefont {Johnson}}, \ and\
  \bibinfo {author} {\bibfnamefont {M.~A.}\ \bibnamefont {Kasevich}},\
  }\bibfield  {title} {\enquote {\bibinfo {title} {Multiaxis inertial sensing
  with long-time point source atom interferometry},}\ }\href {\doibase
  10.1103/PhysRevLett.111.083001} {\bibfield  {journal} {\bibinfo  {journal}
  {Phys. Rev. Lett.}\ }\textbf {\bibinfo {volume} {111}},\ \bibinfo {pages}
  {083001} (\bibinfo {year} {2013})}\BibitemShut {NoStop}%
\bibitem [{\citenamefont {Hoth}\ \emph {et~al.}(2016)\citenamefont {Hoth},
  \citenamefont {Pelle}, \citenamefont {Riedl}, \citenamefont {Kitching},\ and\
  \citenamefont {Donley}}]{Hoth2016}%
  \BibitemOpen
  \bibfield  {author} {\bibinfo {author} {\bibfnamefont {G.~W.}\ \bibnamefont
  {Hoth}}, \bibinfo {author} {\bibfnamefont {B.}~\bibnamefont {Pelle}},
  \bibinfo {author} {\bibfnamefont {S.}~\bibnamefont {Riedl}}, \bibinfo
  {author} {\bibfnamefont {J.}~\bibnamefont {Kitching}}, \ and\ \bibinfo
  {author} {\bibfnamefont {E.~A.}\ \bibnamefont {Donley}},\ }\bibfield  {title}
  {\enquote {\bibinfo {title} {Point source atom interferometry with a cloud of
  finite size},}\ }\href {\doibase 10.1063/1.4961527} {\bibfield  {journal}
  {\bibinfo  {journal} {Appl. Phys. Lett.}\ }\textbf {\bibinfo {volume}
  {109}},\ \bibinfo {pages} {071113} (\bibinfo {year} {2016})}\BibitemShut
  {NoStop}%
\bibitem [{\citenamefont {Chen}\ \emph {et~al.}(2019)\citenamefont {Chen},
  \citenamefont {Hansen}, \citenamefont {Hoth}, \citenamefont {Ivanov},
  \citenamefont {Pelle}, \citenamefont {Kitching},\ and\ \citenamefont
  {Donley}}]{Chen2019}%
  \BibitemOpen
  \bibfield  {author} {\bibinfo {author} {\bibfnamefont {Y.-J.}\ \bibnamefont
  {Chen}}, \bibinfo {author} {\bibfnamefont {A.}~\bibnamefont {Hansen}},
  \bibinfo {author} {\bibfnamefont {G.~W.}\ \bibnamefont {Hoth}}, \bibinfo
  {author} {\bibfnamefont {E.}~\bibnamefont {Ivanov}}, \bibinfo {author}
  {\bibfnamefont {B.}~\bibnamefont {Pelle}}, \bibinfo {author} {\bibfnamefont
  {J.}~\bibnamefont {Kitching}}, \ and\ \bibinfo {author} {\bibfnamefont
  {E.~A.}\ \bibnamefont {Donley}},\ }\bibfield  {title} {\enquote {\bibinfo
  {title} {Single-source multiaxis cold-atom interferometer in a
  centimeter-scale cell},}\ }\href {\doibase 10.1103/PhysRevApplied.12.014019}
  {\bibfield  {journal} {\bibinfo  {journal} {Phys. Rev. Applied}\ }\textbf
  {\bibinfo {volume} {12}},\ \bibinfo {pages} {014019} (\bibinfo {year}
  {2019})}\BibitemShut {NoStop}%
\bibitem [{\citenamefont {Wu}\ \emph {et~al.}(2007)\citenamefont {Wu},
  \citenamefont {Su},\ and\ \citenamefont {Prentiss}}]{Wu2007}%
  \BibitemOpen
  \bibfield  {author} {\bibinfo {author} {\bibfnamefont {S.}~\bibnamefont
  {Wu}}, \bibinfo {author} {\bibfnamefont {E.}~\bibnamefont {Su}}, \ and\
  \bibinfo {author} {\bibfnamefont {M.}~\bibnamefont {Prentiss}},\ }\bibfield
  {title} {\enquote {\bibinfo {title} {Demonstration of an area-enclosing
  guided-atom interferometer for rotation sensing},}\ }\href {\doibase
  10.1103/PhysRevLett.99.173201} {\bibfield  {journal} {\bibinfo  {journal}
  {Phys. Rev. Lett.}\ }\textbf {\bibinfo {volume} {99}},\ \bibinfo {pages}
  {173201} (\bibinfo {year} {2007})}\BibitemShut {NoStop}%
\bibitem [{\citenamefont {Moan}\ \emph {et~al.}(2020)\citenamefont {Moan},
  \citenamefont {Horne}, \citenamefont {Arpornthip}, \citenamefont {Luo},
  \citenamefont {Fallon}, \citenamefont {Berl},\ and\ \citenamefont
  {Sackett}}]{Moan2020}%
  \BibitemOpen
  \bibfield  {author} {\bibinfo {author} {\bibfnamefont {E.~R.}\ \bibnamefont
  {Moan}}, \bibinfo {author} {\bibfnamefont {R.~A.}\ \bibnamefont {Horne}},
  \bibinfo {author} {\bibfnamefont {T.}~\bibnamefont {Arpornthip}}, \bibinfo
  {author} {\bibfnamefont {Z.}~\bibnamefont {Luo}}, \bibinfo {author}
  {\bibfnamefont {A.~J.}\ \bibnamefont {Fallon}}, \bibinfo {author}
  {\bibfnamefont {S.~J.}\ \bibnamefont {Berl}}, \ and\ \bibinfo {author}
  {\bibfnamefont {C.~A.}\ \bibnamefont {Sackett}},\ }\bibfield  {title}
  {\enquote {\bibinfo {title} {Quantum rotation sensing with dual {S}agnac
  interferometers in an atom-optical waveguide},}\ }\href {\doibase
  10.1103/PhysRevLett.124.120403} {\bibfield  {journal} {\bibinfo  {journal}
  {Phys. Rev. Lett.}\ }\textbf {\bibinfo {volume} {124}},\ \bibinfo {pages}
  {120403} (\bibinfo {year} {2020})}\BibitemShut {NoStop}%
\bibitem [{\citenamefont {Sapiro}\ \emph {et~al.}(2009)\citenamefont {Sapiro},
  \citenamefont {Zhang},\ and\ \citenamefont {Raithel}}]{Sapiro2009}%
  \BibitemOpen
  \bibfield  {author} {\bibinfo {author} {\bibfnamefont {R.~E.}\ \bibnamefont
  {Sapiro}}, \bibinfo {author} {\bibfnamefont {R.}~\bibnamefont {Zhang}}, \
  and\ \bibinfo {author} {\bibfnamefont {G.}~\bibnamefont {Raithel}},\
  }\bibfield  {title} {\enquote {\bibinfo {title} {Atom interferometry using
  {K}apitza-{D}irac scattering in a magnetic trap},}\ }\href {\doibase
  10.1103/PhysRevA.79.043630} {\bibfield  {journal} {\bibinfo  {journal} {Phys.
  Rev. A}\ }\textbf {\bibinfo {volume} {79}},\ \bibinfo {pages} {043630}
  (\bibinfo {year} {2009})}\BibitemShut {NoStop}%
\bibitem [{\citenamefont {Davis}\ and\ \citenamefont
  {Narducci}(2008)}]{Davis2008}%
  \BibitemOpen
  \bibfield  {author} {\bibinfo {author} {\bibfnamefont {J.P.}\ \bibnamefont
  {Davis}}\ and\ \bibinfo {author} {\bibfnamefont {F.A.}\ \bibnamefont
  {Narducci}},\ }\bibfield  {title} {\enquote {\bibinfo {title} {A proposal for
  a gradient magnetometer atom interferometer},}\ }\href {\doibase
  10.1080/09500340802468633} {\bibfield  {journal} {\bibinfo  {journal} {J.
  Mod. Opt.}\ }\textbf {\bibinfo {volume} {55}},\ \bibinfo {pages} {3173--3185}
  (\bibinfo {year} {2008})}\BibitemShut {NoStop}%
\bibitem [{\citenamefont {Zimmermann}\ \emph {et~al.}(2019)\citenamefont
  {Zimmermann}, \citenamefont {Efremov}, \citenamefont {Zeller}, \citenamefont
  {Schleich}, \citenamefont {Davis},\ and\ \citenamefont
  {Narducci}}]{Zimmermann2019}%
  \BibitemOpen
  \bibfield  {author} {\bibinfo {author} {\bibfnamefont {M.}~\bibnamefont
  {Zimmermann}}, \bibinfo {author} {\bibfnamefont {M.~A.}\ \bibnamefont
  {Efremov}}, \bibinfo {author} {\bibfnamefont {W.}~\bibnamefont {Zeller}},
  \bibinfo {author} {\bibfnamefont {W.~P.}\ \bibnamefont {Schleich}}, \bibinfo
  {author} {\bibfnamefont {J.~P.}\ \bibnamefont {Davis}}, \ and\ \bibinfo
  {author} {\bibfnamefont {F.~A.}\ \bibnamefont {Narducci}},\ }\bibfield
  {title} {\enquote {\bibinfo {title} {Representation-free description of atom
  interferometers in time-dependent linear potentials},}\ }\href {\doibase
  10.1088/1367-2630/ab2e8c} {\bibfield  {journal} {\bibinfo  {journal} {New J.
  Phys.}\ }\textbf {\bibinfo {volume} {21}},\ \bibinfo {pages} {073031}
  (\bibinfo {year} {2019})}\BibitemShut {NoStop}%
\bibitem [{\citenamefont {Chen}\ \emph {et~al.}(2020)\citenamefont {Chen},
  \citenamefont {Hansen}, \citenamefont {Shuker}, \citenamefont {Boudot},
  \citenamefont {Kitching},\ and\ \citenamefont {Donley}}]{Chen2020}%
  \BibitemOpen
  \bibfield  {author} {\bibinfo {author} {\bibfnamefont {Y.-J.}\ \bibnamefont
  {Chen}}, \bibinfo {author} {\bibfnamefont {A.}~\bibnamefont {Hansen}},
  \bibinfo {author} {\bibfnamefont {M.}~\bibnamefont {Shuker}}, \bibinfo
  {author} {\bibfnamefont {R.}~\bibnamefont {Boudot}}, \bibinfo {author}
  {\bibfnamefont {J.}~\bibnamefont {Kitching}}, \ and\ \bibinfo {author}
  {\bibfnamefont {E.~A.}\ \bibnamefont {Donley}},\ }\bibfield  {title}
  {\enquote {\bibinfo {title} {Robust inertial sensing with point-source atom
  interferometry for interferograms spanning a partial period},}\ }\href
  {\doibase 10.1364/OE.399988} {\bibfield  {journal} {\bibinfo  {journal} {Opt.
  Express}\ }\textbf {\bibinfo {volume} {28}},\ \bibinfo {pages} {34516--34529}
  (\bibinfo {year} {2020})}\BibitemShut {NoStop}%
\bibitem [{\citenamefont {Stickney}\ and\ \citenamefont
  {Zozulya}(2002)}]{Stickney2002}%
  \BibitemOpen
  \bibfield  {author} {\bibinfo {author} {\bibfnamefont {J.~A.}\ \bibnamefont
  {Stickney}}\ and\ \bibinfo {author} {\bibfnamefont {A.~A.}\ \bibnamefont
  {Zozulya}},\ }\bibfield  {title} {\enquote {\bibinfo {title} {Wave-function
  recombination instability in cold-atom interferometers},}\ }\href {\doibase
  10.1103/PhysRevA.66.053601} {\bibfield  {journal} {\bibinfo  {journal} {Phys.
  Rev. A}\ }\textbf {\bibinfo {volume} {66}},\ \bibinfo {pages} {053601}
  (\bibinfo {year} {2002})}\BibitemShut {NoStop}%
\bibitem [{\citenamefont {Stickney}\ and\ \citenamefont
  {Zozulya}(2003)}]{Stickney2003}%
  \BibitemOpen
  \bibfield  {author} {\bibinfo {author} {\bibfnamefont {J.~A.}\ \bibnamefont
  {Stickney}}\ and\ \bibinfo {author} {\bibfnamefont {A.~A.}\ \bibnamefont
  {Zozulya}},\ }\bibfield  {title} {\enquote {\bibinfo {title} {Influence of
  nonadiabaticity and nonlinearity on the operation of cold-atom beam
  splitters},}\ }\href {\doibase 10.1103/PhysRevA.68.013611} {\bibfield
  {journal} {\bibinfo  {journal} {Phys. Rev. A}\ }\textbf {\bibinfo {volume}
  {68}},\ \bibinfo {pages} {013611} (\bibinfo {year} {2003})}\BibitemShut
  {NoStop}%
\bibitem [{\citenamefont {Burke}\ \emph {et~al.}(2008)\citenamefont {Burke},
  \citenamefont {Deissler}, \citenamefont {Hughes},\ and\ \citenamefont
  {Sackett}}]{Burke2008}%
  \BibitemOpen
  \bibfield  {author} {\bibinfo {author} {\bibfnamefont {J.~H.~T.}\
  \bibnamefont {Burke}}, \bibinfo {author} {\bibfnamefont {B.}~\bibnamefont
  {Deissler}}, \bibinfo {author} {\bibfnamefont {K.~J.}\ \bibnamefont
  {Hughes}}, \ and\ \bibinfo {author} {\bibfnamefont {C.~A.}\ \bibnamefont
  {Sackett}},\ }\bibfield  {title} {\enquote {\bibinfo {title} {Confinement
  effects in a guided-wave atom interferometer with millimeter-scale arm
  separation},}\ }\href {\doibase 10.1103/PhysRevA.78.023619} {\bibfield
  {journal} {\bibinfo  {journal} {Phys. Rev. A}\ }\textbf {\bibinfo {volume}
  {78}},\ \bibinfo {pages} {023619} (\bibinfo {year} {2008})}\BibitemShut
  {NoStop}%
\bibitem [{\citenamefont {Stickney}\ \emph {et~al.}(2008)\citenamefont
  {Stickney}, \citenamefont {Kafle}, \citenamefont {Anderson},\ and\
  \citenamefont {Zozulya}}]{Stickney2008}%
  \BibitemOpen
  \bibfield  {author} {\bibinfo {author} {\bibfnamefont {J.~A.}\ \bibnamefont
  {Stickney}}, \bibinfo {author} {\bibfnamefont {R.~P.}\ \bibnamefont {Kafle}},
  \bibinfo {author} {\bibfnamefont {D.~Z.}\ \bibnamefont {Anderson}}, \ and\
  \bibinfo {author} {\bibfnamefont {A.~A.}\ \bibnamefont {Zozulya}},\
  }\bibfield  {title} {\enquote {\bibinfo {title} {Theoretical analysis of a
  single- and double-reflection atom interferometer in a weakly confining
  magnetic trap},}\ }\href {\doibase 10.1103/PhysRevA.77.043604} {\bibfield
  {journal} {\bibinfo  {journal} {Phys. Rev. A}\ }\textbf {\bibinfo {volume}
  {77}},\ \bibinfo {pages} {043604} (\bibinfo {year} {2008})}\BibitemShut
  {NoStop}%
\bibitem [{\citenamefont {et~al.}(2021{\natexlab{a}})}]{Frye2021}%
  \BibitemOpen
  \bibfield  {author} {\bibinfo {author} {\bibfnamefont {K.~Frye}\ \bibnamefont
  {et~al.}},\ }\bibfield  {title} {\enquote {\bibinfo {title} {The
  {B}ose-{E}instein condensate and {C}old {A}tom {L}aboratory},}\ }\href
  {\doibase 10.1140/epjqt/s40507-020-00090-8} {\bibfield  {journal} {\bibinfo
  {journal} {EPJ Quantum Technol.}\ }\textbf {\bibinfo {volume} {8}},\ \bibinfo
  {pages} {1} (\bibinfo {year} {2021}{\natexlab{a}})}\BibitemShut {NoStop}%
\bibitem [{\citenamefont {Duspayev}\ and\ \citenamefont
  {Raithel}(2021)}]{Duspayev2021}%
  \BibitemOpen
  \bibfield  {author} {\bibinfo {author} {\bibfnamefont {A.}~\bibnamefont
  {Duspayev}}\ and\ \bibinfo {author} {\bibfnamefont {G.}~\bibnamefont
  {Raithel}},\ }\bibfield  {title} {\enquote {\bibinfo {title} {Tractor atom
  interferometry},}\ }\href {\doibase 10.1103/PhysRevA.104.013307} {\bibfield
  {journal} {\bibinfo  {journal} {Phys. Rev. A}\ }\textbf {\bibinfo {volume}
  {104}},\ \bibinfo {pages} {013307} (\bibinfo {year} {2021})}\BibitemShut
  {NoStop}%
\bibitem [{\citenamefont {H\"ansel}\ \emph {et~al.}(2001)\citenamefont
  {H\"ansel}, \citenamefont {Reichel}, \citenamefont {Hommelhoff},\ and\
  \citenamefont {H\"ansch}}]{hanselpra}%
  \BibitemOpen
  \bibfield  {author} {\bibinfo {author} {\bibfnamefont {W.}~\bibnamefont
  {H\"ansel}}, \bibinfo {author} {\bibfnamefont {J.}~\bibnamefont {Reichel}},
  \bibinfo {author} {\bibfnamefont {P.}~\bibnamefont {Hommelhoff}}, \ and\
  \bibinfo {author} {\bibfnamefont {T.~W.}\ \bibnamefont {H\"ansch}},\
  }\bibfield  {title} {\enquote {\bibinfo {title} {Trapped-atom interferometer
  in a magnetic microtrap},}\ }\href {\doibase 10.1103/PhysRevA.64.063607}
  {\bibfield  {journal} {\bibinfo  {journal} {Phys. Rev. A}\ }\textbf {\bibinfo
  {volume} {64}},\ \bibinfo {pages} {063607} (\bibinfo {year}
  {2001})}\BibitemShut {NoStop}%
\bibitem [{\citenamefont {Steffen}\ \emph {et~al.}(2012)\citenamefont
  {Steffen}, \citenamefont {Alberti}, \citenamefont {Alt}, \citenamefont
  {Belmechri}, \citenamefont {Hild}, \citenamefont {Karski}, \citenamefont
  {Widera},\ and\ \citenamefont {Meschede}}]{steffen2012}%
  \BibitemOpen
  \bibfield  {author} {\bibinfo {author} {\bibfnamefont {A.}~\bibnamefont
  {Steffen}}, \bibinfo {author} {\bibfnamefont {A.}~\bibnamefont {Alberti}},
  \bibinfo {author} {\bibfnamefont {W.}~\bibnamefont {Alt}}, \bibinfo {author}
  {\bibfnamefont {N.}~\bibnamefont {Belmechri}}, \bibinfo {author}
  {\bibfnamefont {S.}~\bibnamefont {Hild}}, \bibinfo {author} {\bibfnamefont
  {M.}~\bibnamefont {Karski}}, \bibinfo {author} {\bibfnamefont
  {A.}~\bibnamefont {Widera}}, \ and\ \bibinfo {author} {\bibfnamefont
  {D.}~\bibnamefont {Meschede}},\ }\bibfield  {title} {\enquote {\bibinfo
  {title} {Digital atom interferometer with single particle control on a
  discretized space-time geometry},}\ }\href {\doibase 10.1073/pnas.1204285109}
  {\bibfield  {journal} {\bibinfo  {journal} {PNAS}\ }\textbf {\bibinfo
  {volume} {109}},\ \bibinfo {pages} {9770--9774} (\bibinfo {year}
  {2012})}\BibitemShut {NoStop}%
\bibitem [{\citenamefont {Endres}\ \emph {et~al.}(2016)\citenamefont {Endres},
  \citenamefont {Bernien}, \citenamefont {Keesling}, \citenamefont {Levine},
  \citenamefont {Anschuetz}, \citenamefont {Krajenbrink}, \citenamefont
  {Senko}, \citenamefont {Vuletic}, \citenamefont {Greiner},\ and\
  \citenamefont {Lukin}}]{Endres1024}%
  \BibitemOpen
  \bibfield  {author} {\bibinfo {author} {\bibfnamefont {M.}~\bibnamefont
  {Endres}}, \bibinfo {author} {\bibfnamefont {H.}~\bibnamefont {Bernien}},
  \bibinfo {author} {\bibfnamefont {A.}~\bibnamefont {Keesling}}, \bibinfo
  {author} {\bibfnamefont {H.}~\bibnamefont {Levine}}, \bibinfo {author}
  {\bibfnamefont {E.~R.}\ \bibnamefont {Anschuetz}}, \bibinfo {author}
  {\bibfnamefont {A.}~\bibnamefont {Krajenbrink}}, \bibinfo {author}
  {\bibfnamefont {C.}~\bibnamefont {Senko}}, \bibinfo {author} {\bibfnamefont
  {V.}~\bibnamefont {Vuletic}}, \bibinfo {author} {\bibfnamefont
  {M.}~\bibnamefont {Greiner}}, \ and\ \bibinfo {author} {\bibfnamefont
  {M.~D.}\ \bibnamefont {Lukin}},\ }\bibfield  {title} {\enquote {\bibinfo
  {title} {Atom-by-atom assembly of defect-free one-dimensional cold atom
  arrays},}\ }\href {\doibase 10.1126/science.aah3752} {\bibfield  {journal}
  {\bibinfo  {journal} {Science}\ }\textbf {\bibinfo {volume} {354}},\ \bibinfo
  {pages} {1024--1027} (\bibinfo {year} {2016})}\BibitemShut {NoStop}%
\bibitem [{\citenamefont {Barredo}\ \emph {et~al.}(2016)\citenamefont
  {Barredo}, \citenamefont {de~L{\'e}s{\'e}leuc}, \citenamefont {Lienhard},
  \citenamefont {Lahaye},\ and\ \citenamefont {Browaeys}}]{Barredo1021}%
  \BibitemOpen
  \bibfield  {author} {\bibinfo {author} {\bibfnamefont {D.}~\bibnamefont
  {Barredo}}, \bibinfo {author} {\bibfnamefont {S.}~\bibnamefont
  {de~L{\'e}s{\'e}leuc}}, \bibinfo {author} {\bibfnamefont {V.}~\bibnamefont
  {Lienhard}}, \bibinfo {author} {\bibfnamefont {T.}~\bibnamefont {Lahaye}}, \
  and\ \bibinfo {author} {\bibfnamefont {A.}~\bibnamefont {Browaeys}},\
  }\bibfield  {title} {\enquote {\bibinfo {title} {An atom-by-atom assembler of
  defect-free arbitrary two-dimensional atomic arrays},}\ }\href {\doibase
  10.1126/science.aah3778} {\bibfield  {journal} {\bibinfo  {journal}
  {Science}\ }\textbf {\bibinfo {volume} {354}},\ \bibinfo {pages} {1021--1023}
  (\bibinfo {year} {2016})}\BibitemShut {NoStop}%
\bibitem [{\citenamefont {Kim}\ \emph {et~al.}(2016)\citenamefont {Kim},
  \citenamefont {Lee}, \citenamefont {Lee}, \citenamefont {Jo}, \citenamefont
  {Song},\ and\ \citenamefont {Ahn}}]{kim2016}%
  \BibitemOpen
  \bibfield  {author} {\bibinfo {author} {\bibfnamefont {H.}~\bibnamefont
  {Kim}}, \bibinfo {author} {\bibfnamefont {W.}~\bibnamefont {Lee}}, \bibinfo
  {author} {\bibfnamefont {H.}~\bibnamefont {Lee}}, \bibinfo {author}
  {\bibfnamefont {H.}~\bibnamefont {Jo}}, \bibinfo {author} {\bibfnamefont
  {Y.}~\bibnamefont {Song}}, \ and\ \bibinfo {author} {\bibfnamefont
  {J.}~\bibnamefont {Ahn}},\ }\bibfield  {title} {\enquote {\bibinfo {title}
  {In situ single-atom array synthesis using dynamic holographic optical
  tweezers},}\ }\href {\doibase 10.1038/ncomms13317} {\bibfield  {journal}
  {\bibinfo  {journal} {Nat. Comm.}\ }\textbf {\bibinfo {volume} {7}},\
  \bibinfo {pages} {13317} (\bibinfo {year} {2016})}\BibitemShut {NoStop}%
\bibitem [{\citenamefont {Ohl~de Mello}\ \emph {et~al.}(2019)\citenamefont
  {Ohl~de Mello}, \citenamefont {Sch\"affner}, \citenamefont {Werkmann},
  \citenamefont {Preuschoff}, \citenamefont {Kohfahl}, \citenamefont
  {Schlosser},\ and\ \citenamefont {Birkl}}]{ohl}%
  \BibitemOpen
  \bibfield  {author} {\bibinfo {author} {\bibfnamefont {D.}~\bibnamefont
  {Ohl~de Mello}}, \bibinfo {author} {\bibfnamefont {D.}~\bibnamefont
  {Sch\"affner}}, \bibinfo {author} {\bibfnamefont {J.}~\bibnamefont
  {Werkmann}}, \bibinfo {author} {\bibfnamefont {T.}~\bibnamefont
  {Preuschoff}}, \bibinfo {author} {\bibfnamefont {L.}~\bibnamefont {Kohfahl}},
  \bibinfo {author} {\bibfnamefont {M.}~\bibnamefont {Schlosser}}, \ and\
  \bibinfo {author} {\bibfnamefont {G.}~\bibnamefont {Birkl}},\ }\bibfield
  {title} {\enquote {\bibinfo {title} {Defect-free assembly of 2{D} clusters of
  more than 100 single-atom quantum systems},}\ }\href {\doibase
  10.1103/PhysRevLett.122.203601} {\bibfield  {journal} {\bibinfo  {journal}
  {Phys. Rev. Lett.}\ }\textbf {\bibinfo {volume} {122}},\ \bibinfo {pages}
  {203601} (\bibinfo {year} {2019})}\BibitemShut {NoStop}%
\bibitem [{\citenamefont {Kovachy}\ \emph {et~al.}(2010)\citenamefont
  {Kovachy}, \citenamefont {Hogan}, \citenamefont {Johnson},\ and\
  \citenamefont {Kasevich}}]{kovachypra2010}%
  \BibitemOpen
  \bibfield  {author} {\bibinfo {author} {\bibfnamefont {T.}~\bibnamefont
  {Kovachy}}, \bibinfo {author} {\bibfnamefont {J.~M.}\ \bibnamefont {Hogan}},
  \bibinfo {author} {\bibfnamefont {D.~M.~S.}\ \bibnamefont {Johnson}}, \ and\
  \bibinfo {author} {\bibfnamefont {M.~A.}\ \bibnamefont {Kasevich}},\
  }\bibfield  {title} {\enquote {\bibinfo {title} {Optical lattices as
  waveguides and beam splitters for atom interferometry: An analytical
  treatment and proposal of applications},}\ }\href {\doibase
  10.1103/PhysRevA.82.013638} {\bibfield  {journal} {\bibinfo  {journal} {Phys.
  Rev. A}\ }\textbf {\bibinfo {volume} {82}},\ \bibinfo {pages} {013638}
  (\bibinfo {year} {2010})}\BibitemShut {NoStop}%
\bibitem [{\citenamefont {Kumar}\ \emph {et~al.}(2018)\citenamefont {Kumar},
  \citenamefont {Wu}, \citenamefont {Giraldo},\ and\ \citenamefont
  {Weiss}}]{Kumar2018}%
  \BibitemOpen
  \bibfield  {author} {\bibinfo {author} {\bibfnamefont {A.}~\bibnamefont
  {Kumar}}, \bibinfo {author} {\bibfnamefont {T.-Y.}\ \bibnamefont {Wu}},
  \bibinfo {author} {\bibfnamefont {F.}~\bibnamefont {Giraldo}}, \ and\
  \bibinfo {author} {\bibfnamefont {D.~S.}\ \bibnamefont {Weiss}},\ }\bibfield
  {title} {\enquote {\bibinfo {title} {Sorting ultracold atoms in a
  three-dimensional optical lattice in a realization of {M}axwell’s demon},}\
  }\href {\doibase 10.1038/s41586-018-0458-7} {\bibfield  {journal} {\bibinfo
  {journal} {Nature}\ }\textbf {\bibinfo {volume} {561}},\ \bibinfo {pages}
  {83} (\bibinfo {year} {2018})}\BibitemShut {NoStop}%
\bibitem [{\citenamefont {Klostermann}\ \emph {et~al.}(2022)\citenamefont
  {Klostermann}, \citenamefont {Cabrera}, \citenamefont {von Raven},
  \citenamefont {Wienand}, \citenamefont {Schweizer}, \citenamefont {Bloch},\
  and\ \citenamefont {Aidelsburger}}]{klostermann}%
  \BibitemOpen
  \bibfield  {author} {\bibinfo {author} {\bibfnamefont {T.}~\bibnamefont
  {Klostermann}}, \bibinfo {author} {\bibfnamefont {C.~R.}\ \bibnamefont
  {Cabrera}}, \bibinfo {author} {\bibfnamefont {H.}~\bibnamefont {von Raven}},
  \bibinfo {author} {\bibfnamefont {J.~F.}\ \bibnamefont {Wienand}}, \bibinfo
  {author} {\bibfnamefont {C.}~\bibnamefont {Schweizer}}, \bibinfo {author}
  {\bibfnamefont {I.}~\bibnamefont {Bloch}}, \ and\ \bibinfo {author}
  {\bibfnamefont {M.}~\bibnamefont {Aidelsburger}},\ }\bibfield  {title}
  {\enquote {\bibinfo {title} {Fast long-distance transport of cold cesium
  atoms},}\ }\href {\doibase 10.1103/PhysRevA.105.043319} {\bibfield  {journal}
  {\bibinfo  {journal} {Phys. Rev. A}\ }\textbf {\bibinfo {volume} {105}},\
  \bibinfo {pages} {043319} (\bibinfo {year} {2022})}\BibitemShut {NoStop}%
\bibitem [{\citenamefont {Krinner}\ \emph {et~al.}(2018)\citenamefont
  {Krinner}, \citenamefont {Stewart}, \citenamefont {Pazmiño}, \citenamefont
  {Kwon},\ and\ \citenamefont {Schneble}}]{krinner2018}%
  \BibitemOpen
  \bibfield  {author} {\bibinfo {author} {\bibfnamefont {L.}~\bibnamefont
  {Krinner}}, \bibinfo {author} {\bibfnamefont {M.}~\bibnamefont {Stewart}},
  \bibinfo {author} {\bibfnamefont {A.}~\bibnamefont {Pazmiño}}, \bibinfo
  {author} {\bibfnamefont {J.}~\bibnamefont {Kwon}}, \ and\ \bibinfo {author}
  {\bibfnamefont {D.}~\bibnamefont {Schneble}},\ }\bibfield  {title} {\enquote
  {\bibinfo {title} {Spontaneous emission of matter waves from a tunable open
  quantum system},}\ }\href {\doibase 10.1038/s41586-018-0348-z} {\bibfield
  {journal} {\bibinfo  {journal} {Nature}\ }\textbf {\bibinfo {volume} {559}},\
  \bibinfo {pages} {589} (\bibinfo {year} {2018})}\BibitemShut {NoStop}%
\bibitem [{\citenamefont {White}\ \emph {et~al.}(2006)\citenamefont {White},
  \citenamefont {Gao}, \citenamefont {Pasienski},\ and\ \citenamefont
  {DeMarco}}]{white2006}%
  \BibitemOpen
  \bibfield  {author} {\bibinfo {author} {\bibfnamefont {M.}~\bibnamefont
  {White}}, \bibinfo {author} {\bibfnamefont {H.}~\bibnamefont {Gao}}, \bibinfo
  {author} {\bibfnamefont {M.}~\bibnamefont {Pasienski}}, \ and\ \bibinfo
  {author} {\bibfnamefont {B.}~\bibnamefont {DeMarco}},\ }\bibfield  {title}
  {\enquote {\bibinfo {title} {Bose-{E}instein condensates in rf-dressed
  adiabatic potentials},}\ }\href {\doibase 10.1103/PhysRevA.74.023616}
  {\bibfield  {journal} {\bibinfo  {journal} {Phys. Rev. A}\ }\textbf {\bibinfo
  {volume} {74}},\ \bibinfo {pages} {023616} (\bibinfo {year}
  {2006})}\BibitemShut {NoStop}%
\bibitem [{\citenamefont {Colombe}\ \emph {et~al.}(2004)\citenamefont
  {Colombe}, \citenamefont {Knyazchyan}, \citenamefont {Morizot}, \citenamefont
  {Mercier}, \citenamefont {Lorent},\ and\ \citenamefont
  {Perrin}}]{Colombe_2004}%
  \BibitemOpen
  \bibfield  {author} {\bibinfo {author} {\bibfnamefont {Y.}~\bibnamefont
  {Colombe}}, \bibinfo {author} {\bibfnamefont {E.}~\bibnamefont {Knyazchyan}},
  \bibinfo {author} {\bibfnamefont {O.}~\bibnamefont {Morizot}}, \bibinfo
  {author} {\bibfnamefont {B.}~\bibnamefont {Mercier}}, \bibinfo {author}
  {\bibfnamefont {V.}~\bibnamefont {Lorent}}, \ and\ \bibinfo {author}
  {\bibfnamefont {H.}~\bibnamefont {Perrin}},\ }\bibfield  {title} {\enquote
  {\bibinfo {title} {Ultracold atoms confined in rf-induced two-dimensional
  trapping potentials},}\ }\href {\doibase 10.1209/epl/i2004-10095-7}
  {\bibfield  {journal} {\bibinfo  {journal} {Europhys. Lett.}\ }\textbf
  {\bibinfo {volume} {67}},\ \bibinfo {pages} {593--599} (\bibinfo {year}
  {2004})}\BibitemShut {NoStop}%
\bibitem [{\citenamefont {Guarrera}\ \emph {et~al.}(2015)\citenamefont
  {Guarrera}, \citenamefont {Szmuk}, \citenamefont {Reichel},\ and\
  \citenamefont {Rosenbusch}}]{V_Guarrera_2015}%
  \BibitemOpen
  \bibfield  {author} {\bibinfo {author} {\bibfnamefont {V.}~\bibnamefont
  {Guarrera}}, \bibinfo {author} {\bibfnamefont {R.}~\bibnamefont {Szmuk}},
  \bibinfo {author} {\bibfnamefont {J.}~\bibnamefont {Reichel}}, \ and\
  \bibinfo {author} {\bibfnamefont {P.}~\bibnamefont {Rosenbusch}},\ }\bibfield
   {title} {\enquote {\bibinfo {title} {Microwave-dressed state-selective
  potentials for atom interferometry},}\ }\href {\doibase
  10.1088/1367-2630/17/8/083022} {\bibfield  {journal} {\bibinfo  {journal}
  {New J. Phys.}\ }\textbf {\bibinfo {volume} {17}},\ \bibinfo {pages} {083022}
  (\bibinfo {year} {2015})}\BibitemShut {NoStop}%
\bibitem [{\citenamefont {Garraway}\ and\ \citenamefont
  {Perrin}(2016)}]{Garraway_2016}%
  \BibitemOpen
  \bibfield  {author} {\bibinfo {author} {\bibfnamefont {B.~M.}\ \bibnamefont
  {Garraway}}\ and\ \bibinfo {author} {\bibfnamefont {H.}~\bibnamefont
  {Perrin}},\ }\bibfield  {title} {\enquote {\bibinfo {title} {Recent
  developments in trapping and manipulation of atoms with adiabatic
  potentials},}\ }\href {\doibase 10.1088/0953-4075/49/17/172001} {\bibfield
  {journal} {\bibinfo  {journal} {J. Phys. B}\ }\textbf {\bibinfo {volume}
  {49}},\ \bibinfo {pages} {172001} (\bibinfo {year} {2016})}\BibitemShut
  {NoStop}%
\bibitem [{\citenamefont {Navez}\ \emph {et~al.}(2016)\citenamefont {Navez},
  \citenamefont {Pandey}, \citenamefont {Mas}, \citenamefont {Poulios},
  \citenamefont {Fernholz},\ and\ \citenamefont {von Klitzing}}]{Navez_2016}%
  \BibitemOpen
  \bibfield  {author} {\bibinfo {author} {\bibfnamefont {P.}~\bibnamefont
  {Navez}}, \bibinfo {author} {\bibfnamefont {S.}~\bibnamefont {Pandey}},
  \bibinfo {author} {\bibfnamefont {H.}~\bibnamefont {Mas}}, \bibinfo {author}
  {\bibfnamefont {K.}~\bibnamefont {Poulios}}, \bibinfo {author} {\bibfnamefont
  {T.}~\bibnamefont {Fernholz}}, \ and\ \bibinfo {author} {\bibfnamefont
  {W.}~\bibnamefont {von Klitzing}},\ }\bibfield  {title} {\enquote {\bibinfo
  {title} {Matter-wave interferometers using {TAAP} rings},}\ }\href {\doibase
  10.1088/1367-2630/18/7/075014} {\bibfield  {journal} {\bibinfo  {journal}
  {New J. Phys.}\ }\textbf {\bibinfo {volume} {18}},\ \bibinfo {pages} {075014}
  (\bibinfo {year} {2016})}\BibitemShut {NoStop}%
\bibitem [{\citenamefont {Keil}\ \emph {et~al.}(2016)\citenamefont {Keil},
  \citenamefont {Amit}, \citenamefont {Zhou}, \citenamefont {Groswasser},
  \citenamefont {Japha},\ and\ \citenamefont {Folman}}]{keil2016}%
  \BibitemOpen
  \bibfield  {author} {\bibinfo {author} {\bibfnamefont {M.}~\bibnamefont
  {Keil}}, \bibinfo {author} {\bibfnamefont {O.}~\bibnamefont {Amit}}, \bibinfo
  {author} {\bibfnamefont {S.}~\bibnamefont {Zhou}}, \bibinfo {author}
  {\bibfnamefont {D.}~\bibnamefont {Groswasser}}, \bibinfo {author}
  {\bibfnamefont {Y.}~\bibnamefont {Japha}}, \ and\ \bibinfo {author}
  {\bibfnamefont {R.}~\bibnamefont {Folman}},\ }\bibfield  {title} {\enquote
  {\bibinfo {title} {Fifteen years of cold matter on the atom chip: promise,
  realizations, and prospects},}\ }\href {\doibase
  10.1080/09500340.2016.1178820} {\bibfield  {journal} {\bibinfo  {journal} {J.
  Mod. Opt.}\ }\textbf {\bibinfo {volume} {63}},\ \bibinfo {pages} {1840--1885}
  (\bibinfo {year} {2016})}\BibitemShut {NoStop}%
\bibitem [{\citenamefont {Berrada}\ \emph {et~al.}(2013)\citenamefont
  {Berrada}, \citenamefont {van Frank}, \citenamefont {Bücker}, \citenamefont
  {Schumm}, \citenamefont {Schaff},\ and\ \citenamefont
  {Schmiedmayer}}]{berrada}%
  \BibitemOpen
  \bibfield  {author} {\bibinfo {author} {\bibfnamefont {T.}~\bibnamefont
  {Berrada}}, \bibinfo {author} {\bibfnamefont {S.}~\bibnamefont {van Frank}},
  \bibinfo {author} {\bibfnamefont {R.}~\bibnamefont {Bücker}}, \bibinfo
  {author} {\bibfnamefont {T.}~\bibnamefont {Schumm}}, \bibinfo {author}
  {\bibfnamefont {J.-F.}\ \bibnamefont {Schaff}}, \ and\ \bibinfo {author}
  {\bibfnamefont {J.}~\bibnamefont {Schmiedmayer}},\ }\bibfield  {title}
  {\enquote {\bibinfo {title} {Integrated {M}ach–{Z}ehnder interferometer for
  {B}ose–{E}instein condensates},}\ }\href {\doibase 10.1038/ncomms3077}
  {\bibfield  {journal} {\bibinfo  {journal} {Nat. Comm.}\ }\textbf {\bibinfo
  {volume} {4}},\ \bibinfo {pages} {2077} (\bibinfo {year} {2013})}\BibitemShut
  {NoStop}%
\bibitem [{\citenamefont {Berrada}\ \emph {et~al.}(2016)\citenamefont
  {Berrada}, \citenamefont {van Frank}, \citenamefont {B\"ucker}, \citenamefont
  {Schumm}, \citenamefont {Schaff}, \citenamefont {Schmiedmayer}, \citenamefont
  {Jul\'{\i}a-D\'{\i}az},\ and\ \citenamefont {Polls}}]{berrada2016}%
  \BibitemOpen
  \bibfield  {author} {\bibinfo {author} {\bibfnamefont {T.}~\bibnamefont
  {Berrada}}, \bibinfo {author} {\bibfnamefont {S.}~\bibnamefont {van Frank}},
  \bibinfo {author} {\bibfnamefont {R.}~\bibnamefont {B\"ucker}}, \bibinfo
  {author} {\bibfnamefont {T.}~\bibnamefont {Schumm}}, \bibinfo {author}
  {\bibfnamefont {J.-F.}\ \bibnamefont {Schaff}}, \bibinfo {author}
  {\bibfnamefont {J.}~\bibnamefont {Schmiedmayer}}, \bibinfo {author}
  {\bibfnamefont {B.}~\bibnamefont {Jul\'{\i}a-D\'{\i}az}}, \ and\ \bibinfo
  {author} {\bibfnamefont {A.}~\bibnamefont {Polls}},\ }\bibfield  {title}
  {\enquote {\bibinfo {title} {Matter-wave recombiners for trapped
  {B}ose-{E}instein condensates},}\ }\href {\doibase
  10.1103/PhysRevA.93.063620} {\bibfield  {journal} {\bibinfo  {journal} {Phys.
  Rev. A}\ }\textbf {\bibinfo {volume} {93}},\ \bibinfo {pages} {063620}
  (\bibinfo {year} {2016})}\BibitemShut {NoStop}%
\bibitem [{\citenamefont {Bayha}\ \emph {et~al.}(2020)\citenamefont {Bayha},
  \citenamefont {Holten}, \citenamefont {Klemt}, \citenamefont {Subramanian},
  \citenamefont {Bjerlin}, \citenamefont {Reimann}, \citenamefont {Bruun},
  \citenamefont {Preiss},\ and\ \citenamefont {Jochim}}]{bayha2020}%
  \BibitemOpen
  \bibfield  {author} {\bibinfo {author} {\bibfnamefont {L.}~\bibnamefont
  {Bayha}}, \bibinfo {author} {\bibfnamefont {M.}~\bibnamefont {Holten}},
  \bibinfo {author} {\bibfnamefont {R.}~\bibnamefont {Klemt}}, \bibinfo
  {author} {\bibfnamefont {K.}~\bibnamefont {Subramanian}}, \bibinfo {author}
  {\bibfnamefont {J.}~\bibnamefont {Bjerlin}}, \bibinfo {author} {\bibfnamefont
  {S.~M.}\ \bibnamefont {Reimann}}, \bibinfo {author} {\bibfnamefont {G.~M.}\
  \bibnamefont {Bruun}}, \bibinfo {author} {\bibfnamefont {P.~M.}\ \bibnamefont
  {Preiss}}, \ and\ \bibinfo {author} {\bibfnamefont {S.}~\bibnamefont
  {Jochim}},\ }\bibfield  {title} {\enquote {\bibinfo {title} {Observing the
  emergence of a quantum phase transition shell by shell},}\ }\href {\doibase
  0.1038/s41586-020-2936-y} {\bibfield  {journal} {\bibinfo  {journal}
  {Nature}\ }\textbf {\bibinfo {volume} {587}},\ \bibinfo {pages} {583}
  (\bibinfo {year} {2020})}\BibitemShut {NoStop}%
\bibitem [{\citenamefont {Deist}\ \emph {et~al.}(2022)\citenamefont {Deist},
  \citenamefont {Gerber}, \citenamefont {Lu}, \citenamefont {Zeiher},\ and\
  \citenamefont {Stamper-Kurn}}]{deist2022}%
  \BibitemOpen
  \bibfield  {author} {\bibinfo {author} {\bibfnamefont {E.}~\bibnamefont
  {Deist}}, \bibinfo {author} {\bibfnamefont {J.~A.}\ \bibnamefont {Gerber}},
  \bibinfo {author} {\bibfnamefont {Y.-H.}\ \bibnamefont {Lu}}, \bibinfo
  {author} {\bibfnamefont {J.}~\bibnamefont {Zeiher}}, \ and\ \bibinfo {author}
  {\bibfnamefont {D.~M.}\ \bibnamefont {Stamper-Kurn}},\ }\bibfield  {title}
  {\enquote {\bibinfo {title} {Superresolution microscopy of optical fields
  using tweezer-trapped single atoms},}\ }\href {\doibase
  10.1103/PhysRevLett.128.083201} {\bibfield  {journal} {\bibinfo  {journal}
  {Phys. Rev. Lett.}\ }\textbf {\bibinfo {volume} {128}},\ \bibinfo {pages}
  {083201} (\bibinfo {year} {2022})}\BibitemShut {NoStop}%
\bibitem [{\citenamefont {Trisnadi}\ \emph {et~al.}(2022)\citenamefont
  {Trisnadi}, \citenamefont {Zhang}, \citenamefont {Weiss},\ and\ \citenamefont
  {Chin}}]{trisnadi2022}%
  \BibitemOpen
  \bibfield  {author} {\bibinfo {author} {\bibfnamefont {J.}~\bibnamefont
  {Trisnadi}}, \bibinfo {author} {\bibfnamefont {M.}~\bibnamefont {Zhang}},
  \bibinfo {author} {\bibfnamefont {L.}~\bibnamefont {Weiss}}, \ and\ \bibinfo
  {author} {\bibfnamefont {C.}~\bibnamefont {Chin}},\ }\bibfield  {title}
  {\enquote {\bibinfo {title} {Design and construction of a quantum matter
  synthesizer},}\ }\href {https://arxiv.org/abs/2205.10389} {\bibfield
  {journal} {\bibinfo  {journal} {arXiv:2205.10389}\ } (\bibinfo {year}
  {2022})}\BibitemShut {NoStop}%
\bibitem [{\citenamefont {et~al.}(2020{\natexlab{a}})}]{2020SciPy-NMeth}%
  \BibitemOpen
  \bibfield  {author} {\bibinfo {author} {\bibfnamefont {P.~Virtanen}\
  \bibnamefont {et~al.}},\ }\bibfield  {title} {\enquote {\bibinfo {title}
  {{{SciPy} 1.0: Fundamental Algorithms for Scientific Computing in Python}},}\
  }\href {\doibase 10.1038/s41592-019-0686-2} {\bibfield  {journal} {\bibinfo
  {journal} {Nat. Methods}\ }\textbf {\bibinfo {volume} {17}},\ \bibinfo
  {pages} {261--272} (\bibinfo {year} {2020}{\natexlab{a}})}\BibitemShut
  {NoStop}%
\bibitem [{\citenamefont {Sola}\ \emph {et~al.}(2018)\citenamefont {Sola},
  \citenamefont {Chang}, \citenamefont {Malinovskaya},\ and\ \citenamefont
  {Malinovsky}}]{SolaAAMOP2018AsPaper}%
  \BibitemOpen
  \bibfield  {author} {\bibinfo {author} {\bibfnamefont {I.~R.}\ \bibnamefont
  {Sola}}, \bibinfo {author} {\bibfnamefont {B.~Y.}\ \bibnamefont {Chang}},
  \bibinfo {author} {\bibfnamefont {S.~A.}\ \bibnamefont {Malinovskaya}}, \
  and\ \bibinfo {author} {\bibfnamefont {V.~S.}\ \bibnamefont {Malinovsky}},\
  }\bibfield  {title} {\enquote {\bibinfo {title} {Quantum control in
  multilevel systems},}\ }\href {\doibase 10.1016/bs.aamop.2018.02.003}
  {\bibfield  {journal} {\bibinfo  {journal} {Adv. At. Mol. Opt. Phys.}\
  }\textbf {\bibinfo {volume} {67}},\ \bibinfo {pages} {151--256} (\bibinfo
  {year} {2018})}\BibitemShut {NoStop}%
\bibitem [{\citenamefont {Goerz}\ \emph {et~al.}(2022)\citenamefont {Goerz},
  \citenamefont {Carrasco},\ and\ \citenamefont
  {Malinovsky}}]{goerz2022quantum}%
  \BibitemOpen
  \bibfield  {author} {\bibinfo {author} {\bibfnamefont {M.~H}\ \bibnamefont
  {Goerz}}, \bibinfo {author} {\bibfnamefont {S.~C.}\ \bibnamefont {Carrasco}},
  \ and\ \bibinfo {author} {\bibfnamefont {V.~S.}\ \bibnamefont {Malinovsky}},\
  }\bibfield  {title} {\enquote {\bibinfo {title} {Quantum optimal control via
  semi-automatic differentiation},}\ }\href {https://arxiv.org/abs/2205.15044}
  {\bibfield  {journal} {\bibinfo  {journal} {arXiv:2205.15044}\ } (\bibinfo
  {year} {2022})}\BibitemShut {NoStop}%
\bibitem [{\citenamefont {Carrasco}\ \emph {et~al.}(2022)\citenamefont
  {Carrasco}, \citenamefont {Goerz}, \citenamefont {Li}, \citenamefont
  {Colombo}, \citenamefont {Vuleti\'c},\ and\ \citenamefont
  {Malinovsky}}]{PRApp2022}%
  \BibitemOpen
  \bibfield  {author} {\bibinfo {author} {\bibfnamefont {S.~C.}\ \bibnamefont
  {Carrasco}}, \bibinfo {author} {\bibfnamefont {M.~H.}\ \bibnamefont {Goerz}},
  \bibinfo {author} {\bibfnamefont {Z.}~\bibnamefont {Li}}, \bibinfo {author}
  {\bibfnamefont {S.}~\bibnamefont {Colombo}}, \bibinfo {author} {\bibfnamefont
  {V.}~\bibnamefont {Vuleti\'c}}, \ and\ \bibinfo {author} {\bibfnamefont
  {V.~S.}\ \bibnamefont {Malinovsky}},\ }\bibfield  {title} {\enquote {\bibinfo
  {title} {Extreme spin squeezing via optimized one-axis twisting and
  rotations},}\ }\href {\doibase 10.1103/PhysRevApplied.17.064050} {\bibfield
  {journal} {\bibinfo  {journal} {Phys. Rev. Applied}\ }\textbf {\bibinfo
  {volume} {17}},\ \bibinfo {pages} {064050} (\bibinfo {year}
  {2022})}\BibitemShut {NoStop}%
\bibitem [{\citenamefont {Safronova}\ \emph {et~al.}(2018)\citenamefont
  {Safronova}, \citenamefont {Budker}, \citenamefont {DeMille}, \citenamefont
  {Kimball}, \citenamefont {Derevianko},\ and\ \citenamefont
  {Clark}}]{safronovarmp2018}%
  \BibitemOpen
  \bibfield  {author} {\bibinfo {author} {\bibfnamefont {M.~S.}\ \bibnamefont
  {Safronova}}, \bibinfo {author} {\bibfnamefont {D.}~\bibnamefont {Budker}},
  \bibinfo {author} {\bibfnamefont {D.}~\bibnamefont {DeMille}}, \bibinfo
  {author} {\bibfnamefont {D.~F.~Jackson}\ \bibnamefont {Kimball}}, \bibinfo
  {author} {\bibfnamefont {A.}~\bibnamefont {Derevianko}}, \ and\ \bibinfo
  {author} {\bibfnamefont {C.~W.}\ \bibnamefont {Clark}},\ }\bibfield  {title}
  {\enquote {\bibinfo {title} {Search for new physics with atoms and
  molecules},}\ }\href {\doibase 10.1103/RevModPhys.90.025008} {\bibfield
  {journal} {\bibinfo  {journal} {Rev. Mod. Phys.}\ }\textbf {\bibinfo {volume}
  {90}},\ \bibinfo {pages} {025008} (\bibinfo {year} {2018})}\BibitemShut
  {NoStop}%
\bibitem [{\citenamefont {et~al.}(2020{\natexlab{b}})}]{Badurina_2020}%
  \BibitemOpen
  \bibfield  {author} {\bibinfo {author} {\bibfnamefont {L.~Badurina}\
  \bibnamefont {et~al.}},\ }\bibfield  {title} {\enquote {\bibinfo {title}
  {{AION}: an atom interferometer observatory and network},}\ }\href {\doibase
  10.1088/1475-7516/2020/05/011} {\bibfield  {journal} {\bibinfo  {journal} {J.
  Cosmol. Astropart. Phys.}\ }\textbf {\bibinfo {volume} {2020}},\ \bibinfo
  {pages} {011--011} (\bibinfo {year} {2020}{\natexlab{b}})}\BibitemShut
  {NoStop}%
\bibitem [{\citenamefont {Yu}\ and\ \citenamefont {Tinto}(2011)}]{yu2011}%
  \BibitemOpen
  \bibfield  {author} {\bibinfo {author} {\bibfnamefont {N.}~\bibnamefont
  {Yu}}\ and\ \bibinfo {author} {\bibfnamefont {M.}~\bibnamefont {Tinto}},\
  }\bibfield  {title} {\enquote {\bibinfo {title} {Gravitational wave detection
  with single-laser atom interferometers},}\ }\href {\doibase
  10.1007/s10714-010-1055-8} {\bibfield  {journal} {\bibinfo  {journal} {Gen.
  Relativ. Gravit.}\ }\textbf {\bibinfo {volume} {43}},\ \bibinfo {pages}
  {1943} (\bibinfo {year} {2011})}\BibitemShut {NoStop}%
\bibitem [{\citenamefont {et~al.}(2018)}]{canuel2018}%
  \BibitemOpen
  \bibfield  {author} {\bibinfo {author} {\bibfnamefont {B.~Canuel}\
  \bibnamefont {et~al.}},\ }\bibfield  {title} {\enquote {\bibinfo {title}
  {Exploring gravity with the {MIGA} large scale atom interferometer},}\ }\href
  {\doibase 10.1038/s41598-018-32165-z} {\bibfield  {journal} {\bibinfo
  {journal} {Sci. Rep.}\ }\textbf {\bibinfo {volume} {8}},\ \bibinfo {pages}
  {14064} (\bibinfo {year} {2018})}\BibitemShut {NoStop}%
\bibitem [{\citenamefont {et~al.}(2021{\natexlab{b}})}]{Abe_2021}%
  \BibitemOpen
  \bibfield  {author} {\bibinfo {author} {\bibfnamefont {M.~Abe}\ \bibnamefont
  {et~al.}},\ }\bibfield  {title} {\enquote {\bibinfo {title} {Matter-wave
  atomic gradiometer interferometric sensor ({MAGIS}-100)},}\ }\href {\doibase
  10.1088/2058-9565/abf719} {\bibfield  {journal} {\bibinfo  {journal} {Quantum
  Sci. Technol.}\ }\textbf {\bibinfo {volume} {6}},\ \bibinfo {pages} {044003}
  (\bibinfo {year} {2021}{\natexlab{b}})}\BibitemShut {NoStop}%
\end{thebibliography}%
\end{document}